\documentclass[11pt,a4paper]{article}
\usepackage{jcappub}
\pdfoutput=1
\usepackage{color}
\usepackage{graphicx}
\definecolor{AliceBlue}{rgb}{0.94,0.97,1.00}
\definecolor{AntiqueWhite1}{rgb}{1.00,0.94,0.86}
\definecolor{AntiqueWhite2}{rgb}{0.93,0.87,0.80}
\definecolor{AntiqueWhite3}{rgb}{0.80,0.75,0.69}
\definecolor{AntiqueWhite4}{rgb}{0.55,0.51,0.47}
\definecolor{AntiqueWhite}{rgb}{0.98,0.92,0.84}
\definecolor{BlanchedAlmond}{rgb}{1.00,0.92,0.80}
\definecolor{BlueViolet}{rgb}{0.54,0.17,0.89}
\definecolor{CadetBlue1}{rgb}{0.60,0.96,1.00}
\definecolor{CadetBlue2}{rgb}{0.56,0.90,0.93}
\definecolor{CadetBlue3}{rgb}{0.48,0.77,0.80}
\definecolor{CadetBlue4}{rgb}{0.33,0.53,0.55}
\definecolor{CadetBlue}{rgb}{0.37,0.62,0.63}
\definecolor{CornflowerBlue}{rgb}{0.39,0.58,0.93}
\definecolor{DarkBlue}{rgb}{0.00,0.00,0.55}
\definecolor{DarkCyan}{rgb}{0.00,0.55,0.55}
\definecolor{DarkGoldenrod1}{rgb}{1.00,0.73,0.06}
\definecolor{DarkGoldenrod2}{rgb}{0.93,0.68,0.05}
\definecolor{DarkGoldenrod3}{rgb}{0.80,0.58,0.05}
\definecolor{DarkGoldenrod4}{rgb}{0.55,0.40,0.03}
\definecolor{DarkGoldenrod}{rgb}{0.72,0.53,0.04}
\definecolor{DarkGray}{rgb}{0.66,0.66,0.66}
\definecolor{DarkGreen}{rgb}{0.00,0.39,0.00}
\definecolor{DarkGrey}{rgb}{0.66,0.66,0.66}
\definecolor{DarkKhaki}{rgb}{0.74,0.72,0.42}
\definecolor{DarkMagenta}{rgb}{0.55,0.00,0.55}
\definecolor{DarkOliveGreen1}{rgb}{0.79,1.00,0.44}
\definecolor{DarkOliveGreen2}{rgb}{0.74,0.93,0.41}
\definecolor{DarkOliveGreen3}{rgb}{0.64,0.80,0.35}
\definecolor{DarkOliveGreen4}{rgb}{0.43,0.55,0.24}
\definecolor{DarkOliveGreen}{rgb}{0.33,0.42,0.18}
\definecolor{DarkOrange1}{rgb}{1.00,0.50,0.00}
\definecolor{DarkOrange2}{rgb}{0.93,0.46,0.00}
\definecolor{DarkOrange3}{rgb}{0.80,0.40,0.00}
\definecolor{DarkOrange4}{rgb}{0.55,0.27,0.00}
\definecolor{DarkOrange}{rgb}{1.00,0.55,0.00}
\definecolor{DarkOrchid1}{rgb}{0.75,0.24,1.00}
\definecolor{DarkOrchid2}{rgb}{0.70,0.23,0.93}
\definecolor{DarkOrchid3}{rgb}{0.60,0.20,0.80}
\definecolor{DarkOrchid4}{rgb}{0.41,0.13,0.55}
\definecolor{DarkOrchid}{rgb}{0.60,0.20,0.80}
\definecolor{DarkRed}{rgb}{0.55,0.00,0.00}
\definecolor{DarkSalmon}{rgb}{0.91,0.59,0.48}
\definecolor{DarkSeaGreen1}{rgb}{0.76,1.00,0.76}
\definecolor{DarkSeaGreen2}{rgb}{0.71,0.93,0.71}
\definecolor{DarkSeaGreen3}{rgb}{0.61,0.80,0.61}
\definecolor{DarkSeaGreen4}{rgb}{0.41,0.55,0.41}
\definecolor{DarkSeaGreen}{rgb}{0.56,0.74,0.56}
\definecolor{DarkSlateBlue}{rgb}{0.28,0.24,0.55}
\definecolor{DarkSlateGray1}{rgb}{0.59,1.00,1.00}
\definecolor{DarkSlateGray2}{rgb}{0.55,0.93,0.93}
\definecolor{DarkSlateGray3}{rgb}{0.47,0.80,0.80}
\definecolor{DarkSlateGray4}{rgb}{0.32,0.55,0.55}
\definecolor{DarkSlateGray}{rgb}{0.18,0.31,0.31}
\definecolor{DarkSlateGrey}{rgb}{0.18,0.31,0.31}
\definecolor{DarkTurquoise}{rgb}{0.00,0.81,0.82}
\definecolor{DarkViolet}{rgb}{0.58,0.00,0.83}
\definecolor{DeepPink1}{rgb}{1.00,0.08,0.58}
\definecolor{DeepPink2}{rgb}{0.93,0.07,0.54}
\definecolor{DeepPink3}{rgb}{0.80,0.06,0.46}
\definecolor{DeepPink4}{rgb}{0.55,0.04,0.31}
\definecolor{DeepPink}{rgb}{1.00,0.08,0.58}
\definecolor{DeepSkyBlue1}{rgb}{0.00,0.75,1.00}
\definecolor{DeepSkyBlue2}{rgb}{0.00,0.70,0.93}
\definecolor{DeepSkyBlue3}{rgb}{0.00,0.60,0.80}
\definecolor{DeepSkyBlue4}{rgb}{0.00,0.41,0.55}
\definecolor{DeepSkyBlue}{rgb}{0.00,0.75,1.00}
\definecolor{DimGray}{rgb}{0.41,0.41,0.41}
\definecolor{DimGrey}{rgb}{0.41,0.41,0.41}
\definecolor{DodgerBlue1}{rgb}{0.12,0.56,1.00}
\definecolor{DodgerBlue2}{rgb}{0.11,0.53,0.93}
\definecolor{DodgerBlue3}{rgb}{0.09,0.45,0.80}
\definecolor{DodgerBlue4}{rgb}{0.06,0.31,0.55}
\definecolor{DodgerBlue}{rgb}{0.12,0.56,1.00}
\definecolor{FloralWhite}{rgb}{1.00,0.98,0.94}
\definecolor{ForestGreen}{rgb}{0.13,0.55,0.13}
\definecolor{GhostWhite}{rgb}{0.97,0.97,1.00}
\definecolor{GreenYellow}{rgb}{0.68,1.00,0.18}
\definecolor{HotPink1}{rgb}{1.00,0.43,0.71}
\definecolor{HotPink2}{rgb}{0.93,0.42,0.65}
\definecolor{HotPink3}{rgb}{0.80,0.38,0.56}
\definecolor{HotPink4}{rgb}{0.55,0.23,0.38}
\definecolor{HotPink}{rgb}{1.00,0.41,0.71}
\definecolor{IndianRed1}{rgb}{1.00,0.42,0.42}
\definecolor{IndianRed2}{rgb}{0.93,0.39,0.39}
\definecolor{IndianRed3}{rgb}{0.80,0.33,0.33}
\definecolor{IndianRed4}{rgb}{0.55,0.23,0.23}
\definecolor{IndianRed}{rgb}{0.80,0.36,0.36}
\definecolor{LavenderBlush1}{rgb}{1.00,0.94,0.96}
\definecolor{LavenderBlush2}{rgb}{0.93,0.88,0.90}
\definecolor{LavenderBlush3}{rgb}{0.80,0.76,0.77}
\definecolor{LavenderBlush4}{rgb}{0.55,0.51,0.53}
\definecolor{LavenderBlush}{rgb}{1.00,0.94,0.96}
\definecolor{LawnGreen}{rgb}{0.49,0.99,0.00}
\definecolor{LemonChiffon1}{rgb}{1.00,0.98,0.80}
\definecolor{LemonChiffon2}{rgb}{0.93,0.91,0.75}
\definecolor{LemonChiffon3}{rgb}{0.80,0.79,0.65}
\definecolor{LemonChiffon4}{rgb}{0.55,0.54,0.44}
\definecolor{LemonChiffon}{rgb}{1.00,0.98,0.80}
\definecolor{LightBlue1}{rgb}{0.75,0.94,1.00}
\definecolor{LightBlue2}{rgb}{0.70,0.87,0.93}
\definecolor{LightBlue3}{rgb}{0.60,0.75,0.80}
\definecolor{LightBlue4}{rgb}{0.41,0.51,0.55}
\definecolor{LightBlue}{rgb}{0.68,0.85,0.90}
\definecolor{LightCoral}{rgb}{0.94,0.50,0.50}
\definecolor{LightCyan1}{rgb}{0.88,1.00,1.00}
\definecolor{LightCyan2}{rgb}{0.82,0.93,0.93}
\definecolor{LightCyan3}{rgb}{0.71,0.80,0.80}
\definecolor{LightCyan4}{rgb}{0.48,0.55,0.55}
\definecolor{LightCyan}{rgb}{0.88,1.00,1.00}
\definecolor{LightGoldenrod1}{rgb}{1.00,0.93,0.55}
\definecolor{LightGoldenrod2}{rgb}{0.93,0.86,0.51}
\definecolor{LightGoldenrod3}{rgb}{0.80,0.75,0.44}
\definecolor{LightGoldenrod4}{rgb}{0.55,0.51,0.30}
\definecolor{LightGoldenrodYellow}{rgb}{0.98,0.98,0.82}
\definecolor{LightGoldenrod}{rgb}{0.93,0.87,0.51}
\definecolor{LightGray}{rgb}{0.83,0.83,0.83}
\definecolor{LightGreen}{rgb}{0.56,0.93,0.56}
\definecolor{LightGrey}{rgb}{0.83,0.83,0.83}
\definecolor{LightPink1}{rgb}{1.00,0.68,0.73}
\definecolor{LightPink2}{rgb}{0.93,0.64,0.68}
\definecolor{LightPink3}{rgb}{0.80,0.55,0.58}
\definecolor{LightPink4}{rgb}{0.55,0.37,0.40}
\definecolor{LightPink}{rgb}{1.00,0.71,0.76}
\definecolor{LightSalmon1}{rgb}{1.00,0.63,0.48}
\definecolor{LightSalmon2}{rgb}{0.93,0.58,0.45}
\definecolor{LightSalmon3}{rgb}{0.80,0.51,0.38}
\definecolor{LightSalmon4}{rgb}{0.55,0.34,0.26}
\definecolor{LightSalmon}{rgb}{1.00,0.63,0.48}
\definecolor{LightSeaGreen}{rgb}{0.13,0.70,0.67}
\definecolor{LightSkyBlue1}{rgb}{0.69,0.89,1.00}
\definecolor{LightSkyBlue2}{rgb}{0.64,0.83,0.93}
\definecolor{LightSkyBlue3}{rgb}{0.55,0.71,0.80}
\definecolor{LightSkyBlue4}{rgb}{0.38,0.48,0.55}
\definecolor{LightSkyBlue}{rgb}{0.53,0.81,0.98}
\definecolor{LightSlateBlue}{rgb}{0.52,0.44,1.00}
\definecolor{LightSlateGray}{rgb}{0.47,0.53,0.60}
\definecolor{LightSlateGrey}{rgb}{0.47,0.53,0.60}
\definecolor{LightSteelBlue1}{rgb}{0.79,0.88,1.00}
\definecolor{LightSteelBlue2}{rgb}{0.74,0.82,0.93}
\definecolor{LightSteelBlue3}{rgb}{0.64,0.71,0.80}
\definecolor{LightSteelBlue4}{rgb}{0.43,0.48,0.55}
\definecolor{LightSteelBlue}{rgb}{0.69,0.77,0.87}
\definecolor{LightYellow1}{rgb}{1.00,1.00,0.88}
\definecolor{LightYellow2}{rgb}{0.93,0.93,0.82}
\definecolor{LightYellow3}{rgb}{0.80,0.80,0.71}
\definecolor{LightYellow4}{rgb}{0.55,0.55,0.48}
\definecolor{LightYellow}{rgb}{1.00,1.00,0.88}
\definecolor{LimeGreen}{rgb}{0.20,0.80,0.20}
\definecolor{MediumAquamarine}{rgb}{0.40,0.80,0.67}
\definecolor{MediumBlue}{rgb}{0.00,0.00,0.80}
\definecolor{MediumOrchid1}{rgb}{0.88,0.40,1.00}
\definecolor{MediumOrchid2}{rgb}{0.82,0.37,0.93}
\definecolor{MediumOrchid3}{rgb}{0.71,0.32,0.80}
\definecolor{MediumOrchid4}{rgb}{0.48,0.22,0.55}
\definecolor{MediumOrchid}{rgb}{0.73,0.33,0.83}
\definecolor{MediumPurple1}{rgb}{0.67,0.51,1.00}
\definecolor{MediumPurple2}{rgb}{0.62,0.47,0.93}
\definecolor{MediumPurple3}{rgb}{0.54,0.41,0.80}
\definecolor{MediumPurple4}{rgb}{0.36,0.28,0.55}
\definecolor{MediumPurple}{rgb}{0.58,0.44,0.86}
\definecolor{MediumSeaGreen}{rgb}{0.24,0.70,0.44}
\definecolor{MediumSlateBlue}{rgb}{0.48,0.41,0.93}
\definecolor{MediumSpringGreen}{rgb}{0.00,0.98,0.60}
\definecolor{MediumTurquoise}{rgb}{0.28,0.82,0.80}
\definecolor{MediumVioletRed}{rgb}{0.78,0.08,0.52}
\definecolor{MidnightBlue}{rgb}{0.10,0.10,0.44}
\definecolor{MintCream}{rgb}{0.96,1.00,0.98}
\definecolor{MistyRose1}{rgb}{1.00,0.89,0.88}
\definecolor{MistyRose2}{rgb}{0.93,0.84,0.82}
\definecolor{MistyRose3}{rgb}{0.80,0.72,0.71}
\definecolor{MistyRose4}{rgb}{0.55,0.49,0.48}
\definecolor{MistyRose}{rgb}{1.00,0.89,0.88}
\definecolor{NavajoWhite1}{rgb}{1.00,0.87,0.68}
\definecolor{NavajoWhite2}{rgb}{0.93,0.81,0.63}
\definecolor{NavajoWhite3}{rgb}{0.80,0.70,0.55}
\definecolor{NavajoWhite4}{rgb}{0.55,0.47,0.37}
\definecolor{NavajoWhite}{rgb}{1.00,0.87,0.68}
\definecolor{NavyBlue}{rgb}{0.00,0.00,0.50}
\definecolor{OldLace}{rgb}{0.99,0.96,0.90}
\definecolor{OliveDrab1}{rgb}{0.75,1.00,0.24}
\definecolor{OliveDrab2}{rgb}{0.70,0.93,0.23}
\definecolor{OliveDrab3}{rgb}{0.60,0.80,0.20}
\definecolor{OliveDrab4}{rgb}{0.41,0.55,0.13}
\definecolor{OliveDrab}{rgb}{0.42,0.56,0.14}
\definecolor{OrangeRed1}{rgb}{1.00,0.27,0.00}
\definecolor{OrangeRed2}{rgb}{0.93,0.25,0.00}
\definecolor{OrangeRed3}{rgb}{0.80,0.22,0.00}
\definecolor{OrangeRed4}{rgb}{0.55,0.15,0.00}
\definecolor{OrangeRed}{rgb}{1.00,0.27,0.00}
\definecolor{PaleGoldenrod}{rgb}{0.93,0.91,0.67}
\definecolor{PaleGreen1}{rgb}{0.60,1.00,0.60}
\definecolor{PaleGreen2}{rgb}{0.56,0.93,0.56}
\definecolor{PaleGreen3}{rgb}{0.49,0.80,0.49}
\definecolor{PaleGreen4}{rgb}{0.33,0.55,0.33}
\definecolor{PaleGreen}{rgb}{0.60,0.98,0.60}
\definecolor{PaleTurquoise1}{rgb}{0.73,1.00,1.00}
\definecolor{PaleTurquoise2}{rgb}{0.68,0.93,0.93}
\definecolor{PaleTurquoise3}{rgb}{0.59,0.80,0.80}
\definecolor{PaleTurquoise4}{rgb}{0.40,0.55,0.55}
\definecolor{PaleTurquoise}{rgb}{0.69,0.93,0.93}
\definecolor{PaleVioletRed1}{rgb}{1.00,0.51,0.67}
\definecolor{PaleVioletRed2}{rgb}{0.93,0.47,0.62}
\definecolor{PaleVioletRed3}{rgb}{0.80,0.41,0.54}
\definecolor{PaleVioletRed4}{rgb}{0.55,0.28,0.36}
\definecolor{PaleVioletRed}{rgb}{0.86,0.44,0.58}
\definecolor{PapayaWhip}{rgb}{1.00,0.94,0.84}
\definecolor{PeachPuff1}{rgb}{1.00,0.85,0.73}
\definecolor{PeachPuff2}{rgb}{0.93,0.80,0.68}
\definecolor{PeachPuff3}{rgb}{0.80,0.69,0.58}
\definecolor{PeachPuff4}{rgb}{0.55,0.47,0.40}
\definecolor{PeachPuff}{rgb}{1.00,0.85,0.73}
\definecolor{PowderBlue}{rgb}{0.69,0.88,0.90}
\definecolor{RosyBrown1}{rgb}{1.00,0.76,0.76}
\definecolor{RosyBrown2}{rgb}{0.93,0.71,0.71}
\definecolor{RosyBrown3}{rgb}{0.80,0.61,0.61}
\definecolor{RosyBrown4}{rgb}{0.55,0.41,0.41}
\definecolor{RosyBrown}{rgb}{0.74,0.56,0.56}
\definecolor{RoyalBlue1}{rgb}{0.28,0.46,1.00}
\definecolor{RoyalBlue2}{rgb}{0.26,0.43,0.93}
\definecolor{RoyalBlue3}{rgb}{0.23,0.37,0.80}
\definecolor{RoyalBlue4}{rgb}{0.15,0.25,0.55}
\definecolor{RoyalBlue}{rgb}{0.25,0.41,0.88}
\definecolor{SaddleBrown}{rgb}{0.55,0.27,0.07}
\definecolor{SandyBrown}{rgb}{0.96,0.64,0.38}
\definecolor{SeaGreen1}{rgb}{0.33,1.00,0.62}
\definecolor{SeaGreen2}{rgb}{0.31,0.93,0.58}
\definecolor{SeaGreen3}{rgb}{0.26,0.80,0.50}
\definecolor{SeaGreen4}{rgb}{0.18,0.55,0.34}
\definecolor{SeaGreen}{rgb}{0.18,0.55,0.34}
\definecolor{SkyBlue1}{rgb}{0.53,0.81,1.00}
\definecolor{SkyBlue2}{rgb}{0.49,0.75,0.93}
\definecolor{SkyBlue3}{rgb}{0.42,0.65,0.80}
\definecolor{SkyBlue4}{rgb}{0.29,0.44,0.55}
\definecolor{SkyBlue}{rgb}{0.53,0.81,0.92}
\definecolor{SlateBlue1}{rgb}{0.51,0.44,1.00}
\definecolor{SlateBlue2}{rgb}{0.48,0.40,0.93}
\definecolor{SlateBlue3}{rgb}{0.41,0.35,0.80}
\definecolor{SlateBlue4}{rgb}{0.28,0.24,0.55}
\definecolor{SlateBlue}{rgb}{0.42,0.35,0.80}
\definecolor{SlateGray1}{rgb}{0.78,0.89,1.00}
\definecolor{SlateGray2}{rgb}{0.73,0.83,0.93}
\definecolor{SlateGray3}{rgb}{0.62,0.71,0.80}
\definecolor{SlateGray4}{rgb}{0.42,0.48,0.55}
\definecolor{SlateGray}{rgb}{0.44,0.50,0.56}
\definecolor{SlateGrey}{rgb}{0.44,0.50,0.56}
\definecolor{SpringGreen1}{rgb}{0.00,1.00,0.50}
\definecolor{SpringGreen2}{rgb}{0.00,0.93,0.46}
\definecolor{SpringGreen3}{rgb}{0.00,0.80,0.40}
\definecolor{SpringGreen4}{rgb}{0.00,0.55,0.27}
\definecolor{SpringGreen}{rgb}{0.00,1.00,0.50}
\definecolor{SteelBlue1}{rgb}{0.39,0.72,1.00}
\definecolor{SteelBlue2}{rgb}{0.36,0.67,0.93}
\definecolor{SteelBlue3}{rgb}{0.31,0.58,0.80}
\definecolor{SteelBlue4}{rgb}{0.21,0.39,0.55}
\definecolor{SteelBlue}{rgb}{0.27,0.51,0.71}
\definecolor{VioletRed1}{rgb}{1.00,0.24,0.59}
\definecolor{VioletRed2}{rgb}{0.93,0.23,0.55}
\definecolor{VioletRed3}{rgb}{0.80,0.20,0.47}
\definecolor{VioletRed4}{rgb}{0.55,0.13,0.32}
\definecolor{VioletRed}{rgb}{0.82,0.13,0.56}
\definecolor{WhiteSmoke}{rgb}{0.96,0.96,0.96}
\definecolor{YellowGreen}{rgb}{0.60,0.80,0.20}
\definecolor{aliceblue}{rgb}{0.94,0.97,1.00}
\definecolor{antiquewhite}{rgb}{0.98,0.92,0.84}
\definecolor{aquamarine1}{rgb}{0.50,1.00,0.83}
\definecolor{aquamarine2}{rgb}{0.46,0.93,0.78}
\definecolor{aquamarine3}{rgb}{0.40,0.80,0.67}
\definecolor{aquamarine4}{rgb}{0.27,0.55,0.45}
\definecolor{aquamarine}{rgb}{0.50,1.00,0.83}
\definecolor{azure1}{rgb}{0.94,1.00,1.00}
\definecolor{azure2}{rgb}{0.88,0.93,0.93}
\definecolor{azure3}{rgb}{0.76,0.80,0.80}
\definecolor{azure4}{rgb}{0.51,0.55,0.55}
\definecolor{azure}{rgb}{0.94,1.00,1.00}
\definecolor{beige}{rgb}{0.96,0.96,0.86}
\definecolor{bisque1}{rgb}{1.00,0.89,0.77}
\definecolor{bisque2}{rgb}{0.93,0.84,0.72}
\definecolor{bisque3}{rgb}{0.80,0.72,0.62}
\definecolor{bisque4}{rgb}{0.55,0.49,0.42}
\definecolor{bisque}{rgb}{1.00,0.89,0.77}
\definecolor{black}{rgb}{0.00,0.00,0.00}
\definecolor{blanchedalmond}{rgb}{1.00,0.92,0.80}
\definecolor{blue1}{rgb}{0.00,0.00,1.00}
\definecolor{blue2}{rgb}{0.00,0.00,0.93}
\definecolor{blue3}{rgb}{0.00,0.00,0.80}
\definecolor{blue4}{rgb}{0.00,0.00,0.55}
\definecolor{blueviolet}{rgb}{0.54,0.17,0.89}
\definecolor{blue}{rgb}{0.00,0.00,1.00}
\definecolor{brown1}{rgb}{1.00,0.25,0.25}
\definecolor{brown2}{rgb}{0.93,0.23,0.23}
\definecolor{brown3}{rgb}{0.80,0.20,0.20}
\definecolor{brown4}{rgb}{0.55,0.14,0.14}
\definecolor{brown}{rgb}{0.65,0.16,0.16}
\definecolor{burlywood1}{rgb}{1.00,0.83,0.61}
\definecolor{burlywood2}{rgb}{0.93,0.77,0.57}
\definecolor{burlywood3}{rgb}{0.80,0.67,0.49}
\definecolor{burlywood4}{rgb}{0.55,0.45,0.33}
\definecolor{burlywood}{rgb}{0.87,0.72,0.53}
\definecolor{cadetblue}{rgb}{0.37,0.62,0.63}
\definecolor{chartreuse1}{rgb}{0.50,1.00,0.00}
\definecolor{chartreuse2}{rgb}{0.46,0.93,0.00}
\definecolor{chartreuse3}{rgb}{0.40,0.80,0.00}
\definecolor{chartreuse4}{rgb}{0.27,0.55,0.00}
\definecolor{chartreuse}{rgb}{0.50,1.00,0.00}
\definecolor{chocolate1}{rgb}{1.00,0.50,0.14}
\definecolor{chocolate2}{rgb}{0.93,0.46,0.13}
\definecolor{chocolate3}{rgb}{0.80,0.40,0.11}
\definecolor{chocolate4}{rgb}{0.55,0.27,0.07}
\definecolor{chocolate}{rgb}{0.82,0.41,0.12}
\definecolor{coral1}{rgb}{1.00,0.45,0.34}
\definecolor{coral2}{rgb}{0.93,0.42,0.31}
\definecolor{coral3}{rgb}{0.80,0.36,0.27}
\definecolor{coral4}{rgb}{0.55,0.24,0.18}
\definecolor{coral}{rgb}{1.00,0.50,0.31}
\definecolor{cornflowerblue}{rgb}{0.39,0.58,0.93}
\definecolor{cornsilk1}{rgb}{1.00,0.97,0.86}
\definecolor{cornsilk2}{rgb}{0.93,0.91,0.80}
\definecolor{cornsilk3}{rgb}{0.80,0.78,0.69}
\definecolor{cornsilk4}{rgb}{0.55,0.53,0.47}
\definecolor{cornsilk}{rgb}{1.00,0.97,0.86}
\definecolor{cyan1}{rgb}{0.00,1.00,1.00}
\definecolor{cyan2}{rgb}{0.00,0.93,0.93}
\definecolor{cyan3}{rgb}{0.00,0.80,0.80}
\definecolor{cyan4}{rgb}{0.00,0.55,0.55}
\definecolor{cyan}{rgb}{0.00,1.00,1.00}
\definecolor{darkblue}{rgb}{0.00,0.00,0.55}
\definecolor{darkcyan}{rgb}{0.00,0.55,0.55}
\definecolor{darkgoldenrod}{rgb}{0.72,0.53,0.04}
\definecolor{darkgray}{rgb}{0.66,0.66,0.66}
\definecolor{darkgreen}{rgb}{0.00,0.39,0.00}
\definecolor{darkgrey}{rgb}{0.66,0.66,0.66}
\definecolor{darkkhaki}{rgb}{0.74,0.72,0.42}
\definecolor{darkmagenta}{rgb}{0.55,0.00,0.55}
\definecolor{darkolive}{rgb}{0.33,0.42,0.18}
\definecolor{darkorange}{rgb}{1.00,0.55,0.00}
\definecolor{darkorchid}{rgb}{0.60,0.20,0.80}
\definecolor{darkred}{rgb}{0.55,0.00,0.00}
\definecolor{darksalmon}{rgb}{0.91,0.59,0.48}
\definecolor{darksea}{rgb}{0.56,0.74,0.56}
\definecolor{darkslate}{rgb}{0.18,0.31,0.31}
\definecolor{darkslate}{rgb}{0.18,0.31,0.31}
\definecolor{darkslate}{rgb}{0.28,0.24,0.55}
\definecolor{darkturquoise}{rgb}{0.00,0.81,0.82}
\definecolor{darkviolet}{rgb}{0.58,0.00,0.83}
\definecolor{deeppink}{rgb}{1.00,0.08,0.58}
\definecolor{deepsky}{rgb}{0.00,0.75,1.00}
\definecolor{dimgray}{rgb}{0.41,0.41,0.41}
\definecolor{dimgrey}{rgb}{0.41,0.41,0.41}
\definecolor{dodgerblue}{rgb}{0.12,0.56,1.00}
\definecolor{firebrick1}{rgb}{1.00,0.19,0.19}
\definecolor{firebrick2}{rgb}{0.93,0.17,0.17}
\definecolor{firebrick3}{rgb}{0.80,0.15,0.15}
\definecolor{firebrick4}{rgb}{0.55,0.10,0.10}
\definecolor{firebrick}{rgb}{0.70,0.13,0.13}
\definecolor{floralwhite}{rgb}{1.00,0.98,0.94}
\definecolor{forestgreen}{rgb}{0.13,0.55,0.13}
\definecolor{gainsboro}{rgb}{0.86,0.86,0.86}
\definecolor{ghostwhite}{rgb}{0.97,0.97,1.00}
\definecolor{gold1}{rgb}{1.00,0.84,0.00}
\definecolor{gold2}{rgb}{0.93,0.79,0.00}
\definecolor{gold3}{rgb}{0.80,0.68,0.00}
\definecolor{gold4}{rgb}{0.55,0.46,0.00}
\definecolor{goldenrod1}{rgb}{1.00,0.76,0.15}
\definecolor{goldenrod2}{rgb}{0.93,0.71,0.13}
\definecolor{goldenrod3}{rgb}{0.80,0.61,0.11}
\definecolor{goldenrod4}{rgb}{0.55,0.41,0.08}
\definecolor{goldenrod}{rgb}{0.85,0.65,0.13}
\definecolor{gold}{rgb}{1.00,0.84,0.00}
\definecolor{gray0}{rgb}{0.00,0.00,0.00}
\definecolor{gray100}{rgb}{1.00,1.00,1.00}
\definecolor{gray10}{rgb}{0.10,0.10,0.10}
\definecolor{gray11}{rgb}{0.11,0.11,0.11}
\definecolor{gray12}{rgb}{0.12,0.12,0.12}
\definecolor{gray13}{rgb}{0.13,0.13,0.13}
\definecolor{gray14}{rgb}{0.14,0.14,0.14}
\definecolor{gray15}{rgb}{0.15,0.15,0.15}
\definecolor{gray16}{rgb}{0.16,0.16,0.16}
\definecolor{gray17}{rgb}{0.17,0.17,0.17}
\definecolor{gray18}{rgb}{0.18,0.18,0.18}
\definecolor{gray19}{rgb}{0.19,0.19,0.19}
\definecolor{gray1}{rgb}{0.01,0.01,0.01}
\definecolor{gray20}{rgb}{0.20,0.20,0.20}
\definecolor{gray21}{rgb}{0.21,0.21,0.21}
\definecolor{gray22}{rgb}{0.22,0.22,0.22}
\definecolor{gray23}{rgb}{0.23,0.23,0.23}
\definecolor{gray24}{rgb}{0.24,0.24,0.24}
\definecolor{gray25}{rgb}{0.25,0.25,0.25}
\definecolor{gray26}{rgb}{0.26,0.26,0.26}
\definecolor{gray27}{rgb}{0.27,0.27,0.27}
\definecolor{gray28}{rgb}{0.28,0.28,0.28}
\definecolor{gray29}{rgb}{0.29,0.29,0.29}
\definecolor{gray2}{rgb}{0.02,0.02,0.02}
\definecolor{gray30}{rgb}{0.30,0.30,0.30}
\definecolor{gray31}{rgb}{0.31,0.31,0.31}
\definecolor{gray32}{rgb}{0.32,0.32,0.32}
\definecolor{gray33}{rgb}{0.33,0.33,0.33}
\definecolor{gray34}{rgb}{0.34,0.34,0.34}
\definecolor{gray35}{rgb}{0.35,0.35,0.35}
\definecolor{gray36}{rgb}{0.36,0.36,0.36}
\definecolor{gray37}{rgb}{0.37,0.37,0.37}
\definecolor{gray38}{rgb}{0.38,0.38,0.38}
\definecolor{gray39}{rgb}{0.39,0.39,0.39}
\definecolor{gray3}{rgb}{0.03,0.03,0.03}
\definecolor{gray40}{rgb}{0.40,0.40,0.40}
\definecolor{gray41}{rgb}{0.41,0.41,0.41}
\definecolor{gray42}{rgb}{0.42,0.42,0.42}
\definecolor{gray43}{rgb}{0.43,0.43,0.43}
\definecolor{gray44}{rgb}{0.44,0.44,0.44}
\definecolor{gray45}{rgb}{0.45,0.45,0.45}
\definecolor{gray46}{rgb}{0.46,0.46,0.46}
\definecolor{gray47}{rgb}{0.47,0.47,0.47}
\definecolor{gray48}{rgb}{0.48,0.48,0.48}
\definecolor{gray49}{rgb}{0.49,0.49,0.49}
\definecolor{gray4}{rgb}{0.04,0.04,0.04}
\definecolor{gray50}{rgb}{0.50,0.50,0.50}
\definecolor{gray51}{rgb}{0.51,0.51,0.51}
\definecolor{gray52}{rgb}{0.52,0.52,0.52}
\definecolor{gray53}{rgb}{0.53,0.53,0.53}
\definecolor{gray54}{rgb}{0.54,0.54,0.54}
\definecolor{gray55}{rgb}{0.55,0.55,0.55}
\definecolor{gray56}{rgb}{0.56,0.56,0.56}
\definecolor{gray57}{rgb}{0.57,0.57,0.57}
\definecolor{gray58}{rgb}{0.58,0.58,0.58}
\definecolor{gray59}{rgb}{0.59,0.59,0.59}
\definecolor{gray5}{rgb}{0.05,0.05,0.05}
\definecolor{gray60}{rgb}{0.60,0.60,0.60}
\definecolor{gray61}{rgb}{0.61,0.61,0.61}
\definecolor{gray62}{rgb}{0.62,0.62,0.62}
\definecolor{gray63}{rgb}{0.63,0.63,0.63}
\definecolor{gray64}{rgb}{0.64,0.64,0.64}
\definecolor{gray65}{rgb}{0.65,0.65,0.65}
\definecolor{gray66}{rgb}{0.66,0.66,0.66}
\definecolor{gray67}{rgb}{0.67,0.67,0.67}
\definecolor{gray68}{rgb}{0.68,0.68,0.68}
\definecolor{gray69}{rgb}{0.69,0.69,0.69}
\definecolor{gray6}{rgb}{0.06,0.06,0.06}
\definecolor{gray70}{rgb}{0.70,0.70,0.70}
\definecolor{gray71}{rgb}{0.71,0.71,0.71}
\definecolor{gray72}{rgb}{0.72,0.72,0.72}
\definecolor{gray73}{rgb}{0.73,0.73,0.73}
\definecolor{gray74}{rgb}{0.74,0.74,0.74}
\definecolor{gray75}{rgb}{0.75,0.75,0.75}
\definecolor{gray76}{rgb}{0.76,0.76,0.76}
\definecolor{gray77}{rgb}{0.77,0.77,0.77}
\definecolor{gray78}{rgb}{0.78,0.78,0.78}
\definecolor{gray79}{rgb}{0.79,0.79,0.79}
\definecolor{gray7}{rgb}{0.07,0.07,0.07}
\definecolor{gray80}{rgb}{0.80,0.80,0.80}
\definecolor{gray81}{rgb}{0.81,0.81,0.81}
\definecolor{gray82}{rgb}{0.82,0.82,0.82}
\definecolor{gray83}{rgb}{0.83,0.83,0.83}
\definecolor{gray84}{rgb}{0.84,0.84,0.84}
\definecolor{gray85}{rgb}{0.85,0.85,0.85}
\definecolor{gray86}{rgb}{0.86,0.86,0.86}
\definecolor{gray87}{rgb}{0.87,0.87,0.87}
\definecolor{gray88}{rgb}{0.88,0.88,0.88}
\definecolor{gray89}{rgb}{0.89,0.89,0.89}
\definecolor{gray8}{rgb}{0.08,0.08,0.08}
\definecolor{gray90}{rgb}{0.90,0.90,0.90}
\definecolor{gray91}{rgb}{0.91,0.91,0.91}
\definecolor{gray92}{rgb}{0.92,0.92,0.92}
\definecolor{gray93}{rgb}{0.93,0.93,0.93}
\definecolor{gray94}{rgb}{0.94,0.94,0.94}
\definecolor{gray95}{rgb}{0.95,0.95,0.95}
\definecolor{gray96}{rgb}{0.96,0.96,0.96}
\definecolor{gray97}{rgb}{0.97,0.97,0.97}
\definecolor{gray98}{rgb}{0.98,0.98,0.98}
\definecolor{gray99}{rgb}{0.99,0.99,0.99}
\definecolor{gray9}{rgb}{0.09,0.09,0.09}
\definecolor{gray}{rgb}{0.75,0.75,0.75}
\definecolor{green1}{rgb}{0.00,1.00,0.00}
\definecolor{green2}{rgb}{0.00,0.93,0.00}
\definecolor{green3}{rgb}{0.00,0.80,0.00}
\definecolor{green4}{rgb}{0.00,0.55,0.00}
\definecolor{greenyellow}{rgb}{0.68,1.00,0.18}
\definecolor{green}{rgb}{0.00,1.00,0.00}
\definecolor{grey0}{rgb}{0.00,0.00,0.00}
\definecolor{grey100}{rgb}{1.00,1.00,1.00}
\definecolor{grey10}{rgb}{0.10,0.10,0.10}
\definecolor{grey11}{rgb}{0.11,0.11,0.11}
\definecolor{grey12}{rgb}{0.12,0.12,0.12}
\definecolor{grey13}{rgb}{0.13,0.13,0.13}
\definecolor{grey14}{rgb}{0.14,0.14,0.14}
\definecolor{grey15}{rgb}{0.15,0.15,0.15}
\definecolor{grey16}{rgb}{0.16,0.16,0.16}
\definecolor{grey17}{rgb}{0.17,0.17,0.17}
\definecolor{grey18}{rgb}{0.18,0.18,0.18}
\definecolor{grey19}{rgb}{0.19,0.19,0.19}
\definecolor{grey1}{rgb}{0.01,0.01,0.01}
\definecolor{grey20}{rgb}{0.20,0.20,0.20}
\definecolor{grey21}{rgb}{0.21,0.21,0.21}
\definecolor{grey22}{rgb}{0.22,0.22,0.22}
\definecolor{grey23}{rgb}{0.23,0.23,0.23}
\definecolor{grey24}{rgb}{0.24,0.24,0.24}
\definecolor{grey25}{rgb}{0.25,0.25,0.25}
\definecolor{grey26}{rgb}{0.26,0.26,0.26}
\definecolor{grey27}{rgb}{0.27,0.27,0.27}
\definecolor{grey28}{rgb}{0.28,0.28,0.28}
\definecolor{grey29}{rgb}{0.29,0.29,0.29}
\definecolor{grey2}{rgb}{0.02,0.02,0.02}
\definecolor{grey30}{rgb}{0.30,0.30,0.30}
\definecolor{grey31}{rgb}{0.31,0.31,0.31}
\definecolor{grey32}{rgb}{0.32,0.32,0.32}
\definecolor{grey33}{rgb}{0.33,0.33,0.33}
\definecolor{grey34}{rgb}{0.34,0.34,0.34}
\definecolor{grey35}{rgb}{0.35,0.35,0.35}
\definecolor{grey36}{rgb}{0.36,0.36,0.36}
\definecolor{grey37}{rgb}{0.37,0.37,0.37}
\definecolor{grey38}{rgb}{0.38,0.38,0.38}
\definecolor{grey39}{rgb}{0.39,0.39,0.39}
\definecolor{grey3}{rgb}{0.03,0.03,0.03}
\definecolor{grey40}{rgb}{0.40,0.40,0.40}
\definecolor{grey41}{rgb}{0.41,0.41,0.41}
\definecolor{grey42}{rgb}{0.42,0.42,0.42}
\definecolor{grey43}{rgb}{0.43,0.43,0.43}
\definecolor{grey44}{rgb}{0.44,0.44,0.44}
\definecolor{grey45}{rgb}{0.45,0.45,0.45}
\definecolor{grey46}{rgb}{0.46,0.46,0.46}
\definecolor{grey47}{rgb}{0.47,0.47,0.47}
\definecolor{grey48}{rgb}{0.48,0.48,0.48}
\definecolor{grey49}{rgb}{0.49,0.49,0.49}
\definecolor{grey4}{rgb}{0.04,0.04,0.04}
\definecolor{grey50}{rgb}{0.50,0.50,0.50}
\definecolor{grey51}{rgb}{0.51,0.51,0.51}
\definecolor{grey52}{rgb}{0.52,0.52,0.52}
\definecolor{grey53}{rgb}{0.53,0.53,0.53}
\definecolor{grey54}{rgb}{0.54,0.54,0.54}
\definecolor{grey55}{rgb}{0.55,0.55,0.55}
\definecolor{grey56}{rgb}{0.56,0.56,0.56}
\definecolor{grey57}{rgb}{0.57,0.57,0.57}
\definecolor{grey58}{rgb}{0.58,0.58,0.58}
\definecolor{grey59}{rgb}{0.59,0.59,0.59}
\definecolor{grey5}{rgb}{0.05,0.05,0.05}
\definecolor{grey60}{rgb}{0.60,0.60,0.60}
\definecolor{grey61}{rgb}{0.61,0.61,0.61}
\definecolor{grey62}{rgb}{0.62,0.62,0.62}
\definecolor{grey63}{rgb}{0.63,0.63,0.63}
\definecolor{grey64}{rgb}{0.64,0.64,0.64}
\definecolor{grey65}{rgb}{0.65,0.65,0.65}
\definecolor{grey66}{rgb}{0.66,0.66,0.66}
\definecolor{grey67}{rgb}{0.67,0.67,0.67}
\definecolor{grey68}{rgb}{0.68,0.68,0.68}
\definecolor{grey69}{rgb}{0.69,0.69,0.69}
\definecolor{grey6}{rgb}{0.06,0.06,0.06}
\definecolor{grey70}{rgb}{0.70,0.70,0.70}
\definecolor{grey71}{rgb}{0.71,0.71,0.71}
\definecolor{grey72}{rgb}{0.72,0.72,0.72}
\definecolor{grey73}{rgb}{0.73,0.73,0.73}
\definecolor{grey74}{rgb}{0.74,0.74,0.74}
\definecolor{grey75}{rgb}{0.75,0.75,0.75}
\definecolor{grey76}{rgb}{0.76,0.76,0.76}
\definecolor{grey77}{rgb}{0.77,0.77,0.77}
\definecolor{grey78}{rgb}{0.78,0.78,0.78}
\definecolor{grey79}{rgb}{0.79,0.79,0.79}
\definecolor{grey7}{rgb}{0.07,0.07,0.07}
\definecolor{grey80}{rgb}{0.80,0.80,0.80}
\definecolor{grey81}{rgb}{0.81,0.81,0.81}
\definecolor{grey82}{rgb}{0.82,0.82,0.82}
\definecolor{grey83}{rgb}{0.83,0.83,0.83}
\definecolor{grey84}{rgb}{0.84,0.84,0.84}
\definecolor{grey85}{rgb}{0.85,0.85,0.85}
\definecolor{grey86}{rgb}{0.86,0.86,0.86}
\definecolor{grey87}{rgb}{0.87,0.87,0.87}
\definecolor{grey88}{rgb}{0.88,0.88,0.88}
\definecolor{grey89}{rgb}{0.89,0.89,0.89}
\definecolor{grey8}{rgb}{0.08,0.08,0.08}
\definecolor{grey90}{rgb}{0.90,0.90,0.90}
\definecolor{grey91}{rgb}{0.91,0.91,0.91}
\definecolor{grey92}{rgb}{0.92,0.92,0.92}
\definecolor{grey93}{rgb}{0.93,0.93,0.93}
\definecolor{grey94}{rgb}{0.94,0.94,0.94}
\definecolor{grey95}{rgb}{0.95,0.95,0.95}
\definecolor{grey96}{rgb}{0.96,0.96,0.96}
\definecolor{grey97}{rgb}{0.97,0.97,0.97}
\definecolor{grey98}{rgb}{0.98,0.98,0.98}
\definecolor{grey99}{rgb}{0.99,0.99,0.99}
\definecolor{grey9}{rgb}{0.09,0.09,0.09}
\definecolor{grey}{rgb}{0.75,0.75,0.75}
\definecolor{honeydew1}{rgb}{0.94,1.00,0.94}
\definecolor{honeydew2}{rgb}{0.88,0.93,0.88}
\definecolor{honeydew3}{rgb}{0.76,0.80,0.76}
\definecolor{honeydew4}{rgb}{0.51,0.55,0.51}
\definecolor{honeydew}{rgb}{0.94,1.00,0.94}
\definecolor{hotpink}{rgb}{1.00,0.41,0.71}
\definecolor{indianred}{rgb}{0.80,0.36,0.36}
\definecolor{ivory1}{rgb}{1.00,1.00,0.94}
\definecolor{ivory2}{rgb}{0.93,0.93,0.88}
\definecolor{ivory3}{rgb}{0.80,0.80,0.76}
\definecolor{ivory4}{rgb}{0.55,0.55,0.51}
\definecolor{ivory}{rgb}{1.00,1.00,0.94}
\definecolor{khaki1}{rgb}{1.00,0.96,0.56}
\definecolor{khaki2}{rgb}{0.93,0.90,0.52}
\definecolor{khaki3}{rgb}{0.80,0.78,0.45}
\definecolor{khaki4}{rgb}{0.55,0.53,0.31}
\definecolor{khaki}{rgb}{0.94,0.90,0.55}
\definecolor{lavenderblush}{rgb}{1.00,0.94,0.96}
\definecolor{lavender}{rgb}{0.90,0.90,0.98}
\definecolor{lawngreen}{rgb}{0.49,0.99,0.00}
\definecolor{lemonchiffon}{rgb}{1.00,0.98,0.80}
\definecolor{lightblue}{rgb}{0.68,0.85,0.90}
\definecolor{lightcoral}{rgb}{0.94,0.50,0.50}
\definecolor{lightcyan}{rgb}{0.88,1.00,1.00}
\definecolor{lightgoldenrod}{rgb}{0.93,0.87,0.51}
\definecolor{lightgoldenrod}{rgb}{0.98,0.98,0.82}
\definecolor{lightgray}{rgb}{0.83,0.83,0.83}
\definecolor{lightgreen}{rgb}{0.56,0.93,0.56}
\definecolor{lightgrey}{rgb}{0.83,0.83,0.83}
\definecolor{lightpink}{rgb}{1.00,0.71,0.76}
\definecolor{lightsalmon}{rgb}{1.00,0.63,0.48}
\definecolor{lightsea}{rgb}{0.13,0.70,0.67}
\definecolor{lightsky}{rgb}{0.53,0.81,0.98}
\definecolor{lightslate}{rgb}{0.47,0.53,0.60}
\definecolor{lightslate}{rgb}{0.47,0.53,0.60}
\definecolor{lightslate}{rgb}{0.52,0.44,1.00}
\definecolor{lightsteel}{rgb}{0.69,0.77,0.87}
\definecolor{lightyellow}{rgb}{1.00,1.00,0.88}
\definecolor{limegreen}{rgb}{0.20,0.80,0.20}
\definecolor{linen}{rgb}{0.98,0.94,0.90}
\definecolor{magenta1}{rgb}{1.00,0.00,1.00}
\definecolor{magenta2}{rgb}{0.93,0.00,0.93}
\definecolor{magenta3}{rgb}{0.80,0.00,0.80}
\definecolor{magenta4}{rgb}{0.55,0.00,0.55}
\definecolor{magenta}{rgb}{1.00,0.00,1.00}
\definecolor{maroon1}{rgb}{1.00,0.20,0.70}
\definecolor{maroon2}{rgb}{0.93,0.19,0.65}
\definecolor{maroon3}{rgb}{0.80,0.16,0.56}
\definecolor{maroon4}{rgb}{0.55,0.11,0.38}
\definecolor{maroon}{rgb}{0.69,0.19,0.38}
\definecolor{mediumaquamarine}{rgb}{0.40,0.80,0.67}
\definecolor{mediumblue}{rgb}{0.00,0.00,0.80}
\definecolor{mediumorchid}{rgb}{0.73,0.33,0.83}
\definecolor{mediumpurple}{rgb}{0.58,0.44,0.86}
\definecolor{mediumsea}{rgb}{0.24,0.70,0.44}
\definecolor{mediumslate}{rgb}{0.48,0.41,0.93}
\definecolor{mediumspring}{rgb}{0.00,0.98,0.60}
\definecolor{mediumturquoise}{rgb}{0.28,0.82,0.80}
\definecolor{mediumviolet}{rgb}{0.78,0.08,0.52}
\definecolor{midnightblue}{rgb}{0.10,0.10,0.44}
\definecolor{mintcream}{rgb}{0.96,1.00,0.98}
\definecolor{mistyrose}{rgb}{1.00,0.89,0.88}
\definecolor{moccasin}{rgb}{1.00,0.89,0.71}
\definecolor{navajowhite}{rgb}{1.00,0.87,0.68}
\definecolor{navyblue}{rgb}{0.00,0.00,0.50}
\definecolor{navy}{rgb}{0.00,0.00,0.50}
\definecolor{oldlace}{rgb}{0.99,0.96,0.90}
\definecolor{olivedrab}{rgb}{0.42,0.56,0.14}
\definecolor{orange1}{rgb}{1.00,0.65,0.00}
\definecolor{orange2}{rgb}{0.93,0.60,0.00}
\definecolor{orange3}{rgb}{0.80,0.52,0.00}
\definecolor{orange4}{rgb}{0.55,0.35,0.00}
\definecolor{orangered}{rgb}{1.00,0.27,0.00}
\definecolor{orange}{rgb}{1.00,0.65,0.00}
\definecolor{orchid1}{rgb}{1.00,0.51,0.98}
\definecolor{orchid2}{rgb}{0.93,0.48,0.91}
\definecolor{orchid3}{rgb}{0.80,0.41,0.79}
\definecolor{orchid4}{rgb}{0.55,0.28,0.54}
\definecolor{orchid}{rgb}{0.85,0.44,0.84}
\definecolor{palegoldenrod}{rgb}{0.93,0.91,0.67}
\definecolor{palegreen}{rgb}{0.60,0.98,0.60}
\definecolor{paleturquoise}{rgb}{0.69,0.93,0.93}
\definecolor{paleviolet}{rgb}{0.86,0.44,0.58}
\definecolor{papayawhip}{rgb}{1.00,0.94,0.84}
\definecolor{peachpuff}{rgb}{1.00,0.85,0.73}
\definecolor{peru}{rgb}{0.80,0.52,0.25}
\definecolor{pink1}{rgb}{1.00,0.71,0.77}
\definecolor{pink2}{rgb}{0.93,0.66,0.72}
\definecolor{pink3}{rgb}{0.80,0.57,0.62}
\definecolor{pink4}{rgb}{0.55,0.39,0.42}
\definecolor{pink}{rgb}{1.00,0.75,0.80}
\definecolor{plum1}{rgb}{1.00,0.73,1.00}
\definecolor{plum2}{rgb}{0.93,0.68,0.93}
\definecolor{plum3}{rgb}{0.80,0.59,0.80}
\definecolor{plum4}{rgb}{0.55,0.40,0.55}
\definecolor{plum}{rgb}{0.87,0.63,0.87}
\definecolor{powderblue}{rgb}{0.69,0.88,0.90}
\definecolor{purple1}{rgb}{0.61,0.19,1.00}
\definecolor{purple2}{rgb}{0.57,0.17,0.93}
\definecolor{purple3}{rgb}{0.49,0.15,0.80}
\definecolor{purple4}{rgb}{0.33,0.10,0.55}
\definecolor{purple}{rgb}{0.63,0.13,0.94}
\definecolor{red1}{rgb}{1.00,0.00,0.00}
\definecolor{red2}{rgb}{0.93,0.00,0.00}
\definecolor{red3}{rgb}{0.80,0.00,0.00}
\definecolor{red4}{rgb}{0.55,0.00,0.00}
\definecolor{red}{rgb}{1.00,0.00,0.00}
\definecolor{rosybrown}{rgb}{0.74,0.56,0.56}
\definecolor{royalblue}{rgb}{0.25,0.41,0.88}
\definecolor{saddlebrown}{rgb}{0.55,0.27,0.07}
\definecolor{salmon1}{rgb}{1.00,0.55,0.41}
\definecolor{salmon2}{rgb}{0.93,0.51,0.38}
\definecolor{salmon3}{rgb}{0.80,0.44,0.33}
\definecolor{salmon4}{rgb}{0.55,0.30,0.22}
\definecolor{salmon}{rgb}{0.98,0.50,0.45}
\definecolor{sandybrown}{rgb}{0.96,0.64,0.38}
\definecolor{seagreen}{rgb}{0.18,0.55,0.34}
\definecolor{seashell1}{rgb}{1.00,0.96,0.93}
\definecolor{seashell2}{rgb}{0.93,0.90,0.87}
\definecolor{seashell3}{rgb}{0.80,0.77,0.75}
\definecolor{seashell4}{rgb}{0.55,0.53,0.51}
\definecolor{seashell}{rgb}{1.00,0.96,0.93}
\definecolor{sienna1}{rgb}{1.00,0.51,0.28}
\definecolor{sienna2}{rgb}{0.93,0.47,0.26}
\definecolor{sienna3}{rgb}{0.80,0.41,0.22}
\definecolor{sienna4}{rgb}{0.55,0.28,0.15}
\definecolor{sienna}{rgb}{0.63,0.32,0.18}
\definecolor{skyblue}{rgb}{0.53,0.81,0.92}
\definecolor{slateblue}{rgb}{0.42,0.35,0.80}
\definecolor{slategray}{rgb}{0.44,0.50,0.56}
\definecolor{slategrey}{rgb}{0.44,0.50,0.56}
\definecolor{snow1}{rgb}{1.00,0.98,0.98}
\definecolor{snow2}{rgb}{0.93,0.91,0.91}
\definecolor{snow3}{rgb}{0.80,0.79,0.79}
\definecolor{snow4}{rgb}{0.55,0.54,0.54}
\definecolor{snow}{rgb}{1.00,0.98,0.98}
\definecolor{springgreen}{rgb}{0.00,1.00,0.50}
\definecolor{steelblue}{rgb}{0.27,0.51,0.71}
\definecolor{tan1}{rgb}{1.00,0.65,0.31}
\definecolor{tan2}{rgb}{0.93,0.60,0.29}
\definecolor{tan3}{rgb}{0.80,0.52,0.25}
\definecolor{tan4}{rgb}{0.55,0.35,0.17}
\definecolor{tan}{rgb}{0.82,0.71,0.55}
\definecolor{thistle1}{rgb}{1.00,0.88,1.00}
\definecolor{thistle2}{rgb}{0.93,0.82,0.93}
\definecolor{thistle3}{rgb}{0.80,0.71,0.80}
\definecolor{thistle4}{rgb}{0.55,0.48,0.55}
\definecolor{thistle}{rgb}{0.85,0.75,0.85}
\definecolor{tomato1}{rgb}{1.00,0.39,0.28}
\definecolor{tomato2}{rgb}{0.93,0.36,0.26}
\definecolor{tomato3}{rgb}{0.80,0.31,0.22}
\definecolor{tomato4}{rgb}{0.55,0.21,0.15}
\definecolor{tomato}{rgb}{1.00,0.39,0.28}
\definecolor{turquoise1}{rgb}{0.00,0.96,1.00}
\definecolor{turquoise2}{rgb}{0.00,0.90,0.93}
\definecolor{turquoise3}{rgb}{0.00,0.77,0.80}
\definecolor{turquoise4}{rgb}{0.00,0.53,0.55}
\definecolor{turquoise}{rgb}{0.25,0.88,0.82}
\definecolor{violetred}{rgb}{0.82,0.13,0.56}
\definecolor{violet}{rgb}{0.93,0.51,0.93}
\definecolor{wheat1}{rgb}{1.00,0.91,0.73}
\definecolor{wheat2}{rgb}{0.93,0.85,0.68}
\definecolor{wheat3}{rgb}{0.80,0.73,0.59}
\definecolor{wheat4}{rgb}{0.55,0.49,0.40}
\definecolor{wheat}{rgb}{0.96,0.87,0.70}
\definecolor{whitesmoke}{rgb}{0.96,0.96,0.96}
\definecolor{white}{rgb}{1.00,1.00,1.00}
\definecolor{yellow1}{rgb}{1.00,1.00,0.00}
\definecolor{yellow2}{rgb}{0.93,0.93,0.00}
\definecolor{yellow3}{rgb}{0.80,0.80,0.00}
\definecolor{yellow4}{rgb}{0.55,0.55,0.00}
\definecolor{yellowgreen}{rgb}{0.60,0.80,0.20}
\definecolor{yellow}{rgb}{1.00,1.00,0.00}

\title{Neutrino signals from electroweak bremsstrahlung in solar WIMP
  annihilation}

\author[a]{Nicole F. Bell,} 
\author[a]{Amelia J. Brennan}
\author[a,b]{and Thomas D. Jacques}

\affiliation[a]{ARC Centre of Excellence for Particle Physics at the
  Terascale \\ School of Physics, The University of Melbourne,
  Victoria 3010, Australia}

\affiliation[b]{Department of Physics and School of Earth and Space
  Exploration, Arizona State University, Tempe, AZ 85287-1404, USA}

\emailAdd{n.bell@unimelb.edu.au}
\emailAdd{a.brennan@pgrad.unimelb.edu.au}
\emailAdd{thomas.jacques@asu.edu}

\abstract{Bremsstrahlung of $W$ and $Z$ gauge bosons, or photons, can
  be an important dark matter annihilation channel.  In many popular
  models in which the annihilation to a pair of light fermions is
  helicity suppressed, these bremsstrahlung processes can lift the
  suppression and thus become the dominant annihilation channels.  The
  resulting dark matter annihilation products contain a large,
  energetic, neutrino component.  We consider solar WIMP annihilation
  in the case where electroweak bremsstrahlung dominates, and
  calculate the resulting neutrino spectra.  The flux consists of
  primary neutrinos produced in processes such as $\chi\chi\rightarrow
  \bar{\nu}\nu Z$ and $\chi\chi\rightarrow \bar{\nu}\ell W$, and
  secondary neutrinos produced via the decays of gauge bosons and
  charged leptons.  After dealing with the neutrino propagation and
  flavour evolution in the Sun, we consider the prospects for
  detection in neutrino experiments on Earth.  We compare our signal
  with that for annihilation to $W^+W^-$, and show that, for a given
  annihilation rate, the bremsstrahlung annihilation channel produces
  a larger signal by a factor of a few.}

\begin{document}

\maketitle

\section{Introduction}

High-energy neutrinos produced by dark matter (DM) self-annihilation
in the Sun offer an appealing signal for indirect DM detection
\cite{silk1985photino,press1985capture,freese1986can,krauss1986solar,gaisser1986limits,gould1987resonant,srednicki1987high,kamionkowski1991energetic}. See,
e.g.,
refs.~\cite{cirelli2005spectra,blennow2008neutrinos,barger2002indirect,Lehnert:2010vb,erkoca2009muon,belotsky2009muon,barger2010fermion}
for recent work on this topic.  DM particles which pass through the
Sun can scatter, lose energy, and become captured by the
gravitational field of the Sun.  They accumulate in the Sun's core and
eventually reach a density where self-annihilation becomes important.
Most Standard Model (SM) particles produced in the annihilations
interact in the Sun and are hence absorbed, while neutrinos will
escape to be detected in experiments on Earth.  The neutrinos may be
produced directly in the DM annihilations, or as secondaries in the
decays of other SM particles.  A high energy neutrino signal from the
Sun would thus be a compelling indication of dark matter.

The strength of this signal is controlled by the DM-nucleon scattering
cross section which controls the DM capture rate and, if in
equilibrium, the annihilation rate.  Existing limits have been placed
by Super-Kamiokande \cite{abe2011indirect}, IceCube
\cite{abbasi2010limits,abbasi2009limits,deyoung2011particle}, and
other experiments.  The strength of these limits depends on the
annihilation channel assumed for dark matter, with limits for soft
annihilation channels such as $\chi\chi\rightarrow \bar{b} b$ being
weaker than those for hard channels such as $\chi\chi \rightarrow
W^+W^-$.  A comprehensive study of the neutrino spectra for the main
annihilation channels with 2-body final states was performed in
ref. \cite{cirelli2005spectra}.  In the present article we study DM
annihilation processes with 3-body final states, namely bremsstrahlung
processes in which a $W$ or $Z$~\cite{bell2011wrev,bell2011w} or
$\gamma$~\cite{bergstrom1989radiative,flores1989radiative,baltz2003detection,bringmann2008new,bergstrom2008new,barger2009generic}
is radiated.  In certain models, these bremsstrahlung processes have been
shown to be the dominant DM annihilation mode.

For non-relativistic dark matter, it is useful to parametrise the
annihilation cross section as
\begin{equation}
\label{eq:crosssectionparam}
\sigma_A v  = a + bv^2 + {\cal O}(v^4),
\end{equation}
where $v$ is the dark matter relative velocity (in units of $c$). The $a$ term
arises only from $s$-wave annihilation, and the $L^{th}$ partial wave
contribution is suppressed by a factor $v^{2L}$.  For highly
non-relativistic DM with $ v \ll 1$, $s$-wave annihilations, if
permitted, are thus expected to dominate.  In the galactic halo, $v^2
\sim 10^{-6}$; $v$ is even smaller for the DM accumulated in the solar
core.  However, in many Majorana DM models, $s$-wave annihilation to a
pair of fermions,
$\chi \chi \rightarrow f \bar{f}$, 
is \emph{helicity} suppressed by a factor of $(m_f / M_{\chi})^2$ due
to a mismatch between the net helicity and chirality of the fermion
pair~\cite{goldberg1983constraint,krauss1983new}.  A well known example
is the annihilation of neutralino dark matter in supersymmetric
models.

The circumstances in which annihilation to a pair of fermions is
velocity and/or helicity suppressed were extensively detailed in
ref.~\cite{bell2011w}.  It is well known that radiation of a photon is
able to lift this suppression via the electromagnetic (EM)
bremsstrahlung process $\chi \chi \rightarrow \bar{f}f
\gamma$~\cite{bergstrom1989radiative,flores1989radiative,baltz2003detection,bringmann2008new,bergstrom2008new,barger2009generic}.
However, as shown in refs.~\cite{bell2011w,bell2011wrev}, electroweak
(EW) bremsstrahlung is also able to lift this helicity suppression,
such that the $\chi \chi \rightarrow \bar{f}f Z$, $\chi \chi
\rightarrow f \nu W$, and $\chi \chi \rightarrow \bar{f}f \gamma$
annihilation channels all dominate over $\chi \chi \rightarrow
\bar{f}f$.  Physically, the suppression is lifted because the emitted
gauge boson carries away a unit of angular momentum, allowing a
fermion spin-flip, so there is no longer a mismatch between the
chirality of the leptons and their allowed 2-particle spin state.
Relative to the 2-body process, the rate for the 3-body final state
suffers a suppression factor from the radiating particle given by
$\alpha / \pi$.  However, this may be more than compensated for by the
consequent \emph{removal} of the $(m_f/M_{\chi})^2$ suppression
factor.

The importance of these radiative corrections to the dark matter
annihilation process have been examined in a number of recent
papers~\cite{bell2011dark,bell2008electroweak,kachelriess2009role,barger2012bremsstrahlung,ciafaloni1999sudakov,ciafaloni2010tev,ciafaloni2011weak,ciafaloni2011importance,ciafaloni2011initial,garny2011antiproton,garny2012dark,ciafaloni2012electroweak},
with a focus on the indirect detection of dark matter due to
annihilations in the Galactic halo.  Interestingly, a possible hint of
an EM bremsstrahlung signal was found in recent Fermi
data~\cite{Bringmann:2012vr}.  In the present paper we consider the
indirect detection of dark matter via annihilation in the Sun, noting
that the EW bremsstrahlung annihilation channels produce large,
energetic, neutrino fluxes.

This paper aims to produce the final neutrino energy spectra, as they
would be detected on Earth, for solar WIMP annihilation via the EW
bremsstrahlung process, and to compare the resulting signal with that
for other commonly assumed annihilation channels.  In section 2 we
describe the neutrino spectra at production, including primary
neutrinos produced directly in the annihilations, and secondaries
produced from the decay of emitted gauge bosons. In section 3 we
consider the scattering and flavour evolution of the neutrinos as they
propagate through the Sun, to determine the final spectra, while in
section 4 we compare the event rates for bremsstrahlung annihilation
with those for other DM annihilation modes.  Our results are
summarised briefly in section 5.


\section{Neutrino production}

During the process of capture in the Sun, DM particles scatter
elastically with solar nuclei, losing enough energy to become
gravitationally bound.  The capture rate, $C^\odot$, is proportional to
the DM/solar nucleon scattering cross section and depends upon
astrophysical parameters such as the local DM velocity and density.
Once captured, subsequent scattering causes the
DM to become concentrated in the solar core, in a region of size $ \sim 0.01
R_{\odot} \sqrt{100 \,{\mathrm{GeV}}/M_{\chi}}$, where $R_{\odot}$ is
the solar radius; the region is sufficiently small that we may assume all
DM annihilations occur in the centre of the Sun~\cite{cirelli2005spectra}.

The differential neutrino flux is given by
\begin{equation}
\frac{dN_{\nu}}{dE} = 
\frac{\Gamma_{\mathrm{ann}}}{4 \pi d^2} \sum_k {\rm BR}_k \frac{dN_k}{dE}, 
\label{eq:diffflux}
\end{equation}
where $d$ is the distance from the centre of the Sun to the detector,
and $BR_k$ is the branching ratio to a specific final state, $k$.  The
annihilation rate $\Gamma_{\mathrm{ann}}$ depends on the capture rate
$C^\odot$ as
\begin{equation}
\label{eq:annrate}
\Gamma_{\mathrm{ann}} = \frac{1}{2} C^\odot \tanh^2 (t_0 / \tau_A),
\end{equation}
where $t_0$ = 4.5 Gyr is the age of the Sun and $\tau_A$ is the
time-scale for competing capture and annihilation
processes~\cite{griest1987cosmic, gould1987resonant, bottino2002does,
  lundberg2004weakly,cirelli2005spectra}.  This time scale is given by
$\tau_A = 1/\sqrt{C^\odot A^\odot}$, where $A^\odot=\langle \sigma
v\rangle/V_{\mathrm{eff}}$ is the annihilation cross section times the
relative WIMP velocity per volume (see, e.g. the reviews in \cite{bertone2005particle,jungman1996supersymmetric} for
explicit expressions for these rates). This time-scale is often much
smaller than $t_0$, such that an equilibrium is reached between the
capture and annihilation processes with $\Gamma_{\text{ann}} \simeq
C^\odot/2$.  In this case, it is the DM-nucleon scattering cross
section, rather than the annihilation cross section, which determines
the annihilation rate.  We shall return to the question of
capture-annihilation equilibrium in section \ref{parameters}.


\subsection{Dark matter coupling}

We use the leptophilic Majorana DM coupling of ref.~\cite{cao2009dark} as
a simple example of a scenario in which EW bremsstrahlung dominates
the annihilation.  The DM is a gauge-singlet Majorana fermion which
annihilates to leptons via the interaction term
\begin{equation}
\label{eq:Caomodel}
f(\nu \, \, \, l^-)_L \, \varepsilon \, \left ( 
\begin{matrix}
\eta^+ \\
\eta^0
\end{matrix}
\right ) \, \chi + h.c. = f(\nu_L \eta^0 - l_L \eta^+) \chi + h.c. \, \, ,
\end{equation}
where $f$ is a coupling constant, $\varepsilon$ is the $2 \times 2$
SU(2) invariant antisymmetric matrix, and $(\eta^+, \, \eta^0)$ form
a new $SU(2)$ doublet scalar which mediates the annihilation.  An
identical coupling arises in supersymmetric models if we take $\chi$
to be a (bino-like) neutralino and $\eta$ a sfermion doublet.

Ref.~\cite{cao2009dark} proposed this model as an explanation of the
sharp excess in the cosmic ray $e^+ / (e^- + e^+)$ fraction at
energies beyond $\sim$10 GeV, as recently measured by the PAMELA
experiment~\cite{adriani2009anomalous,adriani2010pamela,adriani2009new}.
Here we adopt this model as a simple prototype of the situation in
which bremsstrahlung processes dominate the DM annihilation.
As explained in~\cite{bell2011w,bell2011wrev}, the $s$-wave
contributions to the lowest order annihilation processes
$\chi\chi\rightarrow \bar{\nu}\nu, e^+e^-$ are suppressed by factors
of $(m_l/M_\chi)^2$.  Bremsstrahlung lifts this suppression to become the 
dominant annihilation process.

We shall consider three possibilities for flavour structure of the
branching ratios: DM annihilates to all three lepton flavours equally
($f_e:f_\mu:f_\tau=1:1:1$), to electron-flavoured leptons only
($f_e:f_\mu:f_\tau=1:0:0$), or to $\mu$ and $\tau$ flavours equally
($f_e:f_\mu:f_\tau=0:1:1$).  However, we shall see that neutrino
mixing reduces the sensitivity of the final results to the flavour
structure of the couplings.


\subsection{Neutrino spectra at production}

We determine the energy spectra for the neutrinos at production by
adding the contributions from the primary neutrinos produced directly
in DM annihilation, and the secondary neutrino produced by decays of
gauge bosons or charged leptons.  Throughout, we neglect neutrinos
with energies less than $\sim$ 0.5 GeV.  Note that IceCube, for
example, has a detection threshold of $\sim$ 100 GeV
\cite{halzen2009indirect}, which is reduced to $\sim$ 10 GeV with the
addition of DeepCore \cite{abbasi2012design}.  We shall be
particularly interested in the higher-energy regions of the spectra,
as the bremsstrahlung process leads to neutrino spectra with a
distinct high energy peak; moreover, detection cross sections grow
(approximately linearly) with energy.


\subsubsection{Primary neutrino spectra}

We first consider the primary neutrinos, those produced from the direct DM annihilation channels:
\begin{equation}
\begin{aligned}
\label{eq:bremcases}
\chi \chi & \rightarrow && l^- \nu_l W^+ \, \, , \\
\chi \chi & \rightarrow && l^+ \bar{\nu}_l W^- \, \, , \\
\chi \chi & \rightarrow && \nu_l \bar{\nu}_l Z \, \, .\\
\end{aligned}
\end{equation}
The spectra of primary neutrinos from these processes are given in
\cite{bell2011wrev}, along with calculations for the relative
cross sections.  The contribution from each bremsstrahlung channel is
included according to these relative cross sections; for example the
branching ratio for emission of a $W^-$ is given by $\sigma_{e \nu W}
/ \sigma_{\text{total}}$, where $\sigma_{\text{total}}$ is found by
adding all the bremsstrahlung components together (including the
contributions from $\chi \chi \rightarrow l^+ l^- Z$ and $\chi \chi
\rightarrow l^+ l^- \gamma$, which do not produce primary neutrinos).
Annihilation directly to neutrinos via the suppressed process $\chi
\chi \rightarrow \nu \bar{\nu}$ would produce a monoenergetic spectrum
with $E_\nu=M_\chi$; we neglect the contribution of this subdominant
process in our analysis.


\begin{figure}[t]
\begin{center} 
\vspace*{-1cm}
\hspace{-1.5cm}
\includegraphics[scale=0.6]{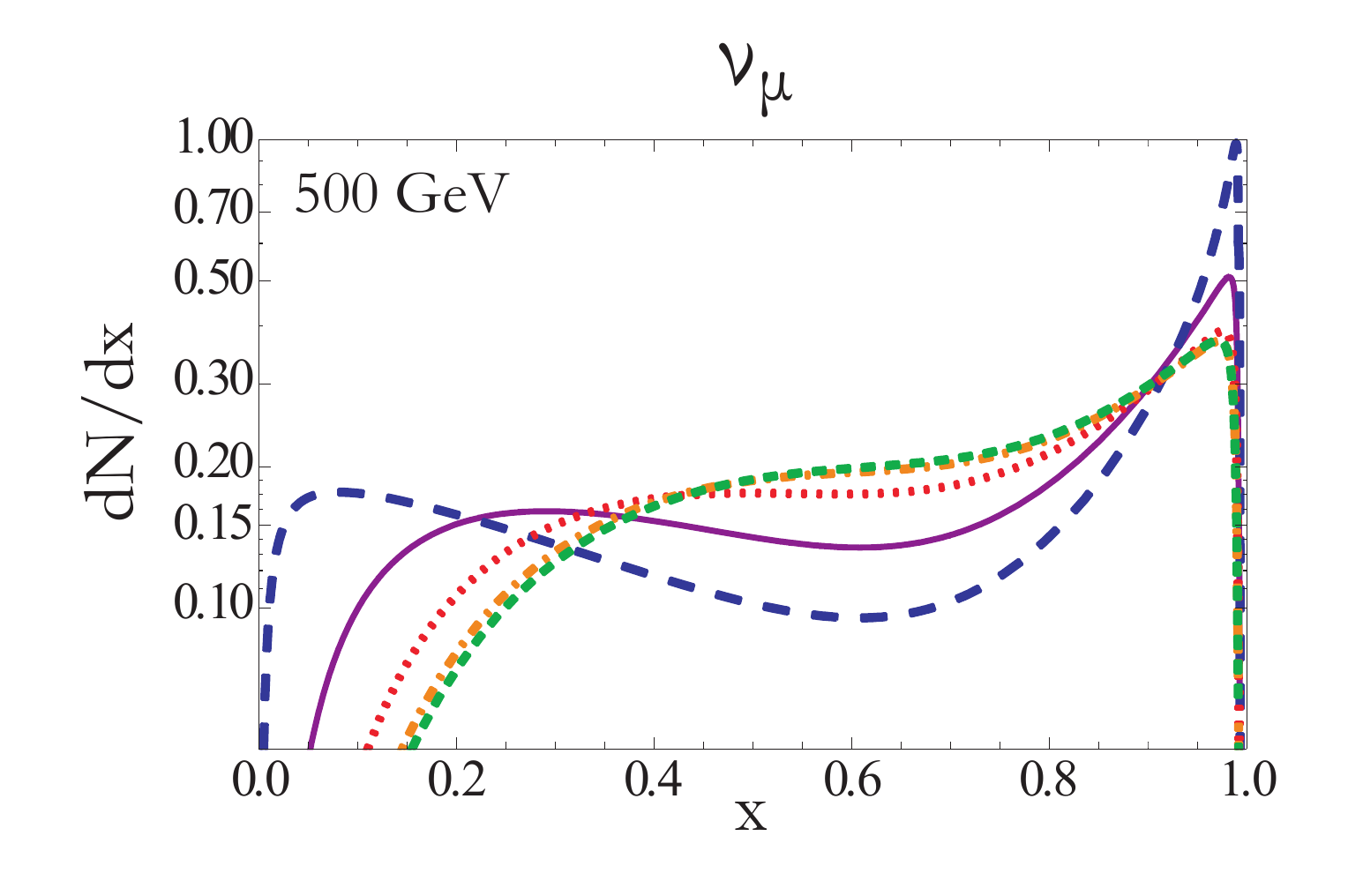}
\caption{The primary spectra of muon-flavoured neutrinos for $\mu = M_{\eta}^2 /
  M_{\chi}^2 =$ 1.01 (dashed, blue), $\mu =$ 1.2 (solid, purple), $\mu
  =$ 2 (red, dotted), $\mu =$ 5 (orange, dot-dashed) and $\mu =$ 10
  (green, small-dashed). Increasing $\mu$ reduces the high-energy peak
  but increases the number of mid-range energy neutrinos.  We have
  taken the annihilations to have equal branching ratios to all lepton
  flavours, $f_e=f_\mu=f_\tau$. Note that the $x$-axis is defined as $x \equiv E_{\nu}/M_{\chi}$.}
\label{fig:VaryingMu}
\end{center}
\end{figure}


The primary neutrino spectrum is sensitive to the ratio of the DM and
propagator masses, $\mu = \frac{M_{\eta}^2}{M_{\chi}^2}$.  We focus on
small values of $\mu$ as this maximises the branching ratio of the
bremsstrahlung processes relative to the 2-body final state
annihilation processes; however, bremsstrahlung remains the dominant
annihilation channel for $\mu \lesssim 10$. In figure \ref{fig:VaryingMu}
we plot the primary neutrino spectra for various choices of $\mu$.
The spectra have a broad energy distribution, and a distinctive high
energy peak at $x\simeq 1$.  Increasing $\mu$ decreases the height of
this peak, but increases the number of neutrinos of mid-range energy. For the remainder of this paper we shall assume a value of $\mu = 1.2$ (but will also show results for $\mu = $5 in table 1).

\begin{figure}[t]
\vspace*{-1cm}
\hspace{-0.5cm}
\includegraphics[scale=0.77]{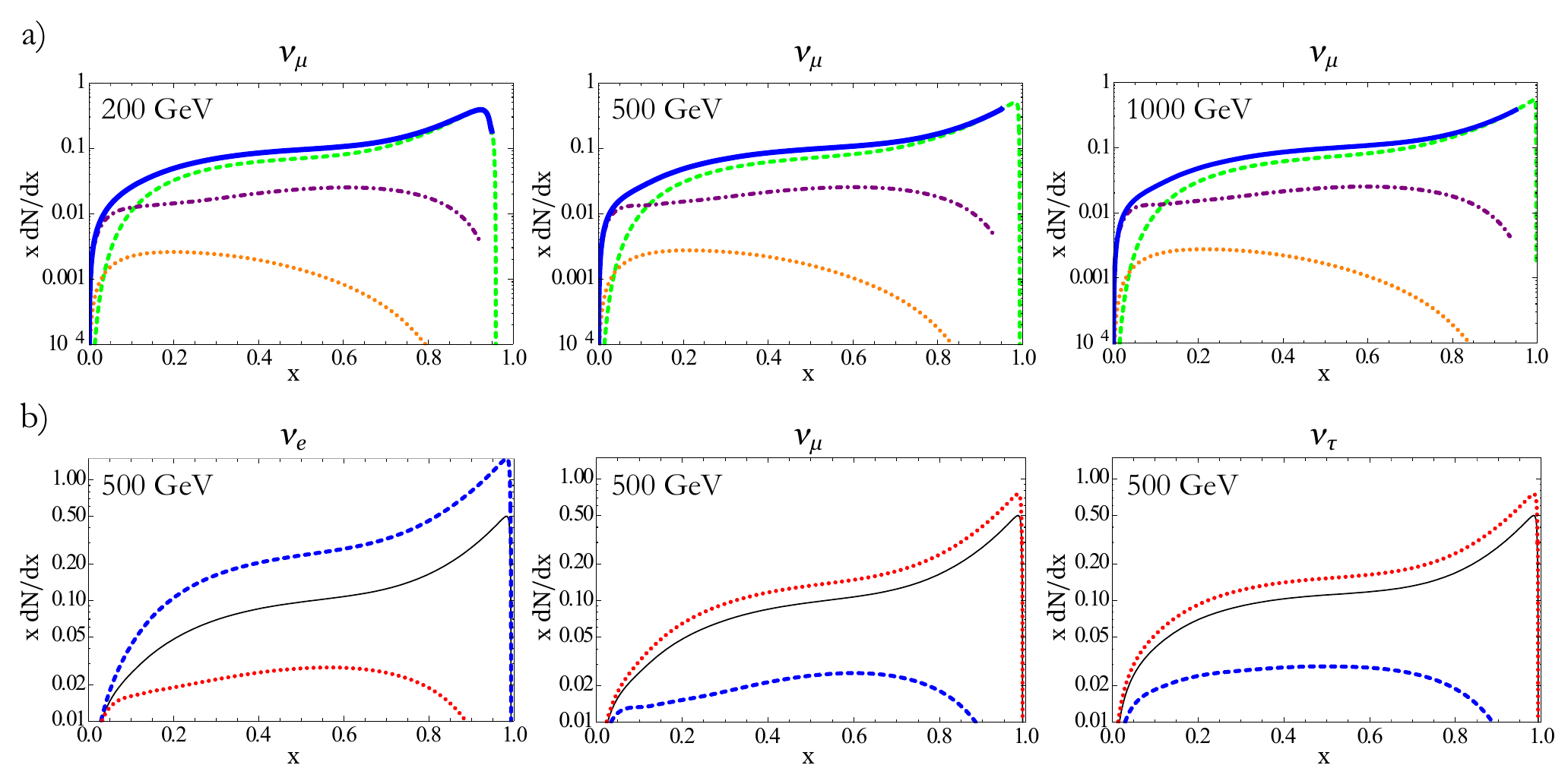}
\caption{a) The contributions to the injected spectra, assuming branching to all lepton flavours equally, from primary neutrinos (green, dashed), secondary neutrinos from gauge boson decay (purple, dot-dashed) and secondary neutrinos from decay of primary taus (orange, dotted); the solid blue line shows the total. b) The total injected spectra for different branching ratios: annihilation to all lepton flavours equally (black, solid), annihilation to $e$-flavoured leptons only (blue, dashed), and annihilation to $\mu$ and $\tau$ flavours equally (red, dotted). All anti-neutrino spectra are identical to the corresponding neutrino spectra, and so are not shown. Note that the $x$-axis is defined as $x \equiv E_{\nu}/M_{\chi}$, and that we are using $\mu =$ 1.2.}
\label{fig:ContrPlus}
\end{figure}

In figure \ref{fig:ContrPlus}(a) the primary spectra of muon neutrinos
are indicated by the dashed green line.  Here we assume equal
branching to all lepton flavours, and so the primary spectra are
flavour-independent; changing the branching ratios will clearly affect
the flavour dependence.  
Note that the primary neutrinos are the dominant contribution to the total neutrino spectrum for each flavour (shown as the solid blue line in figure \ref{fig:ContrPlus}(a)), particularly for the high-energy region of interest.
Comparing the spectra for different choices
of the DM mass, we see that the primary spectra maintain the same
general shape.  However, increasing the DM mass relative to $M_{W/Z}$
decreases the effect of kinematic thresholds and moves the peak closer
to $x=1$.  The antineutrino spectra are identical to the corresponding
neutrino spectra.

\subsubsection{Secondary neutrino spectra}

We next consider the secondary neutrinos that are produced from gauge
boson decays, or the decays of primary tau leptons. The decay of the emitted
gauge boson is modelled using the Monte Carlo simulator
PYTHIA~\cite{sjostrand2008brief}, in which we create a $W$ or $Z$ at
rest which is then allowed to decay with constraints to be outlined
below. The neutrino spectra from the gauge boson decays are then
boosted into the lab frame as per the appendix of
ref.~\cite{bell2011w}.
In the Sun, the core density is $\rho = 140$ g $\mathrm{cm}^{-3}$, so
energy loss processes are relatively important as the DM annihilation
products may interact with solar material and lose energy before
decaying further. Adopting a procedure similar to that in
ref.~\cite{cirelli2005spectra}, the relevant energy losses are
accounted for as described below.

{\it Gauge bosons:} Once produced, the lifetimes of the $W$ and $Z$
bosons are short enough that they decay before interacting and losing
any significant amount of energy. Therefore, in PYTHIA, we set these
bosons to decay as if in vacuum. The products of the hadronic decay
modes of the gauge bosons are dealt with as explained below.
	
{\it Charged leptons:} The typical stopping time of a charged lepton
is $\tau_{\mathrm{stop}} \sim 10^{-11}$s. If this is shorter than the
boosted lifetime, $\tau_{\mathrm{stop}} < \gamma \tau_{\mathrm{dec}}$,
the lepton will lose all its energy before decaying.  For the energy
range of interest, $E \lesssim$ 1 TeV, the muons are stopped before
they decay while taus decay without losing a significant amount of
energy.  Thus, in the PYTHIA routine, we discard electrons and muons,
and allow taus to decay as if in vacuum.

{\it Quarks:} Heavy hadrons have comparable $\tau_{\mathrm{stop}}$ and
$\gamma \tau_{\mathrm{dec}}$, of order $10^{-11}$s and $10^{-12}$s
respectively, so will often lose some energy before decaying; light
hadrons are usually stopped before decay. As a reasonable, simple,
approximation, we allow heavy hadrons (those including charm, top and
bottom quarks) to decay, while light hadrons are discarded.

The resulting neutrino spectra from $W^{\pm}$ and $Z$ decay, as seen
in the lab frame, are shown by the dot-dashed purple line in figure
\ref{fig:ContrPlus}(a). These secondary neutrinos pile up and become
dominant at low energies, as they are often produced through a chain
of decays starting at the gauge boson. For this case of even branching
ratios ($f_e=f_\mu=f_\tau$) the secondary spectra of $\nu_e$ and
$\nu_{\mu}$ are equal, as light hadrons and muons are stopped before
decay, while $\tau$, $c$ and $b$ produce equal numbers of each. As
with the primary neutrinos, the antineutrino spectra are identical to
their corresponding neutrino spectra.

In the case where the coupling $f_\tau$ is non-zero, there will be a
primary tau component which will decay to contribute to the secondary
neutrino flux.  As with the gauge bosons decays, the secondary
neutrino spectra from tau decay (again created using PYTHIA) must be
boosted into the lab frame and convolved with the primary tau
spectrum. The neutrino spectra resulting from this process are shown
by the dotted orange line in figure \ref{fig:ContrPlus}(a).  The tau
decays make a minor contribution to the $\nu_e$ and $\nu_\mu$ fluxes
and a larger, though still subdominant, contribution to the $\nu_\tau$
flux.

Adding the contributions of primary and secondary neutrino fluxes, we
plot the total spectra at production as the solid blue line in figure
\ref{fig:ContrPlus}(a). The total spectra are also shown in figure
\ref{fig:ContrPlus}(b) for varied branching ratios.  We note that the
primary neutrinos make the dominant contribution to the total neutrino
flux (summed over all flavours), while for particular choices of the
flavour branching ratios the secondary neutrinos can dominate the flux
for an individual flavour.  However, once neutrino mixing is taken
into account the primary neutrinos will make the dominant contribution
to the flux for each flavour.  Therefore, our results are robust
against uncertainties in the calculation of the secondary neutrinos,
and instead are determined mainly by the flux of primary neutrinos
(whose spectra can be calculated analytically).


\section{Propagation in the Sun}

Following production, the neutrino flux is modified by flavour
oscillations, and by scattering with solar nuclei, which results in
some absorption of the flux at high energy.  These effects are
well-understood, and various techniques exist to deal with them; we
shall follow the method of ref.~\cite{bell2011enhanced} to account for
the scattering interaction, and then adopt results of
ref.~\cite{friedland2001evolution} when evaluating the flavour
evolution.  Many authors combine scattering and mixing effects into a
single 3$\times$3 density matrix equation that describes the neutrino
evolution.  However in the energy range of interest, $E_{\nu}
\lesssim$ 1 TeV, oscillations and scattering approximately
decouple~\cite{bell2011enhanced}, as the matter effects in the Sun
ensure negligible mixing between $\nu_e$ and $\nu_{\mu,\tau}$ until
\emph{after} scattering becomes unimportant.
We assume throughout that the $\nu_{\mu}$ and $\nu_{\tau}$ fluxes are
equal, as $\nu_{\mu}$ and $\nu_{\tau}$ mix maximally (to good
approximation) and, for the energy range of interest, the oscillation
length is less than or approximately equal to the interaction
length~\cite{bell2011enhanced}.


\subsection{Scattering and absorption}

The evolution equation, used to evolve the neutrino flux from production to $r_c$ (the radius where interactions become unimportant), has the form
\begin{equation}
\label{eq:evolutioneqn}
\frac{\partial \rho_l}{\partial r} = \frac{\partial \rho_l}{\partial r} \biggl\lvert_{\mathrm{inj}} + \frac{\partial \rho_l}{\partial r} \biggl\lvert_{\mathrm{NC}} + \frac{\partial \rho_l}{\partial r} \biggl\lvert_{\mathrm{CC}},
\end{equation}
for neutrinos of flavour $l$, where $\rho(r,E)dE$ is the neutrino flux. The first term on the right corresponds to the neutrino injection due to the annihilation of DM (including all secondary neutrinos). The second and third terms correspond to neutrino scattering through neutral current and charged current interactions respectively.

The probability of an interaction between a neutrino and a solar
nucleon is given by
\begin{equation}
\label{eq:probofint}
P_{\mathrm{int}} = \int_0^{R_\odot} \! \sigma_{\mathrm{tot}} (E) N_S(r') dr'
 = \frac{E}{\mathcal{E}},
\end{equation}
where $\sigma_{\mathrm{tot}}$ is the neutrino total interaction cross section and $N_S(r)\simeq N_0 e^{-r/\kappa R_\odot}$ is the nucleon density of the Sun,
with $\kappa=0.1$ and $N_0 = 1.3 \times 10^{26}$ $\mathrm{cm}^{-3}$.
The constants $\mathcal{E}$ parametrise the energy scale at which
interactions become significant, and are given by
\begin{equation}
\label{eq:curlyEvalues}
\mathcal{E}_{\nu} \simeq 140 \text{ GeV}, \, \, \, \, \, \, \, \, \, \mathcal{E}_{\bar{\nu}} \simeq 213 \text{ GeV}.
\end{equation}

Note that the total interaction cross section combines neutral current (NC) scatterings  $ \nu_l N \leftrightarrow \nu_l N$ and $ \bar{\nu}_l N \leftrightarrow \bar{\nu}_l N$ (effectively removing a neutrino from the flux and re-injecting it at lower energy), and charged current (CC) scatterings $ \nu_l N \leftrightarrow l^- N'$ and $ \bar{\nu}_l N \leftrightarrow l^+ N'$ (effectively removing a neutrino from the flux and producing an almost collinear charged lepton). If $l = e,\mu$ then the final lepton loses most, if not all, of its energy before it can decay again, and so the neutrino is considered absorbed. If $l = \tau$, the resulting $\tau^{\pm}$ decays promptly, either leptonically or hadronically.

Defining the optical depth as
\begin{equation}
\label{eq:opticaldepth}
y(r) \equiv \frac{1}{N_0 \kappa R_{\odot}} \int_0^r N_S (r') dr',
\end{equation}
the evolution equation (\ref{eq:evolutioneqn}) takes the form
\begin{equation}
\label{eq:evoeqn2}
\frac{\partial \rho_j}{\partial y} = h(E) + \frac{1}{\mathcal{E}} \bigg [ - E \, \rho_j(y,E) + \int_E^{\infty} dE' \, g_j(E/E') \, \rho_j(y,E) \bigg] \, \, .
\end{equation}
The term $h(E)$ represents our injected flux, the second term accounts
for absorption of the flux through scattering, and the last term the
reinjection or regeneration of neutrinos at lower energy.
At high energies, absorption through scattering is a significant
effect.  Tau regeneration is also important, as absorbed neutrinos are
repopulated and pile-up at lower energies.  Again, this effect is most
significant for large DM mass.  The regeneration functions $g_i$ are
given, e.g., in
refs.~\cite{cirelli2005spectra,bell2011enhanced}. Following
ref.~\cite{bell2011enhanced} we use a perturbative solution of
eq.~(\ref{eq:evoeqn2}) to handle the regeneration. For simplicity, we
take $g_i = 0.5$ and have checked that for the largest $M_\chi$ we
consider (where regeneration is expected to occur most strongly) this
simplification produces an error $<$20$\%$ in the final spectra for
energies $\gtrsim$ 100 GeV, which does not qualitatively change our
conclusions.


\subsection{Flavour oscillations}

We now include the effect of neutrino flavour mixing. We use the
neutrino mixing parameters given in ref.~\cite{nakamura2010review},
and assume the normal mass hierarchy, where $m_1 < m_2 < m_3$.  We
shall adopt a non-zero value for $\theta_{13}$, as indicated by
recents results from the T2K~\cite{abe2011indication} and the Daya
Bay~\cite{an2012observation} collaborations, and demonstrate how the
results change as we vary the value of this parameter.

Neutrinos undergo flavour conversion during their propagation through
the Sun. These flavour effects differ from oscillations in vacuum due
to the presence of neutrino refractive indices.  In the dense solar
medium, the electron neutrinos experience a matter potential of
$V_e(r) = \sqrt{2} G_F N_e(r)$, where $N_e$ is the electron number
density.  At high density, this leads to an approximate alignment
between the flavour and effective mass eigenstates, and suppresses
oscillations of $\nu_e$ with $\nu_{\mu,\tau}$.  The matter potential
decreases as the neutrinos propagate outward from the centre of the
Sun, and the neutrinos undergo an MSW
resonance~\cite{wolfenstein1978neutrino, mikheev1985resonance} when
\begin{equation}
\label{eq:res}
\Delta m^2/2E \simeq V_e.
\end{equation}  
At low energies, the propagation through this resonance is adiabatic
and the neutrinos remain in given mass eigenstates.  At higher
energies the evolution becomes non-adiabatic, with the neutrinos
experiencing a level-crossing, or jump between mass eigenstates.
Assuming an exponentially varying solar profile ($\rho(r) \propto
e^{-r/r_0}$), with $r_0 = R_\odot/10.54$, the crossing probability is
conventionally given by~\cite{toshev1987exact}
\begin{equation}
\label{eq:Pc}
P_C \, = \frac{e^{\gamma \mathrm{cos}^2 \theta}-1}{e^{\gamma}-1}, \, \, \gamma = 4 \pi r_0 \Delta \, \, \, \, ,
\end{equation}
where $\Delta \equiv \Delta m^2/4E_{\nu}$.  For antineutrinos, the
$\bar{\nu}_e$ potential is $-V_e(r)$, which is equivalent to making
the replacement $\theta \rightarrow \frac{\pi}{2} - \theta$
\cite{strumia2006neutrino}, thus
\begin{equation}
\label{eq:Pcbar}
\bar{P}_C = P_C ( \theta \rightarrow \frac{\pi}{2} - \theta ) \, \, \, .
\end{equation} 

We shall be concerned with two level crossings, controlled by $\Delta
m^2_{21}$ and $\Delta m^2_{32}$ respectively.  As $\Delta m_{32}^2$ is
almost two orders of magnitude larger than $\Delta m_{21}^2$ the
crossings decouple and can be treated independently. This separation
is useful as it allows us to use a simple two-neutrino formalism which
we apply to each crossing independently.  This leads to
\begin{equation}
\begin{aligned}
\label{eq:totalPc}
P &=&& 
 \begin{pmatrix}
  1 - P_C^l & P_C^l & 0 \\
  P_C^l & 1 - P_C^l & 0 \\
  0 & 0 & 1 \\
 \end{pmatrix}
 \begin{pmatrix}
  1 & 0 & 0 \\
  0 &  1 - P_C^h & P_C^h \\
  0 & P_C^h & 1 - P_C^h \\
 \end{pmatrix},\\
\end{aligned} 
\end{equation}
where $P_{jk}$ represents the probability of a neutrino that
originated in matter mass eigenstate $k$ ending up in \emph{vacuum}
mass eigenstate $j$. Here, $P_C^l$ and $P_C^h$ correspond to the lower
and higher level crossing probabilities, driven by $\theta_{12}$ and
$\theta_{13}$ respectively.  See ref.~\cite{lehnert2008neutrino} for a
detailed discussion.

We shall need to evaluate the level crossing probabilities for the two
resonances, and for both neutrinos and antineutrinos.  In some of
these cases the resonance condition~(\ref{eq:res}) is never actually
met; nonetheless, the level crossing probability is non-zero and the
expression given in eq.(\ref{eq:Pc}) still
holds~\cite{mikheev1987resonance,friedland2001evolution,kachelriess2001nonadiabatic}.\footnote{As
  discussed in
  refs.~\cite{kachelriess2001nonadiabatic,friedland2001evolution}, the
  crossing probability is largest when adiabaticity is maximally
  violated, and is non-zero even away from resonance.}  For some
parameters of interest (namely, high neutrino energy) the lower level
crossing takes place in the convective zone of the Sun where the solar
density profile deviates from the exponential form assumed in
eq. (\ref{eq:Pc}).  We account for this by following the procedure
developed in ref.~\cite{friedland2001evolution}, where $P_C$ is
calculated according to eq.~(\ref{eq:Pc}), but with $r_0$ replaced by
an energy-dependent numerical fit.

\begin{figure}
\vspace*{-1cm}
\hspace{-1cm} 
\includegraphics[scale=0.83]{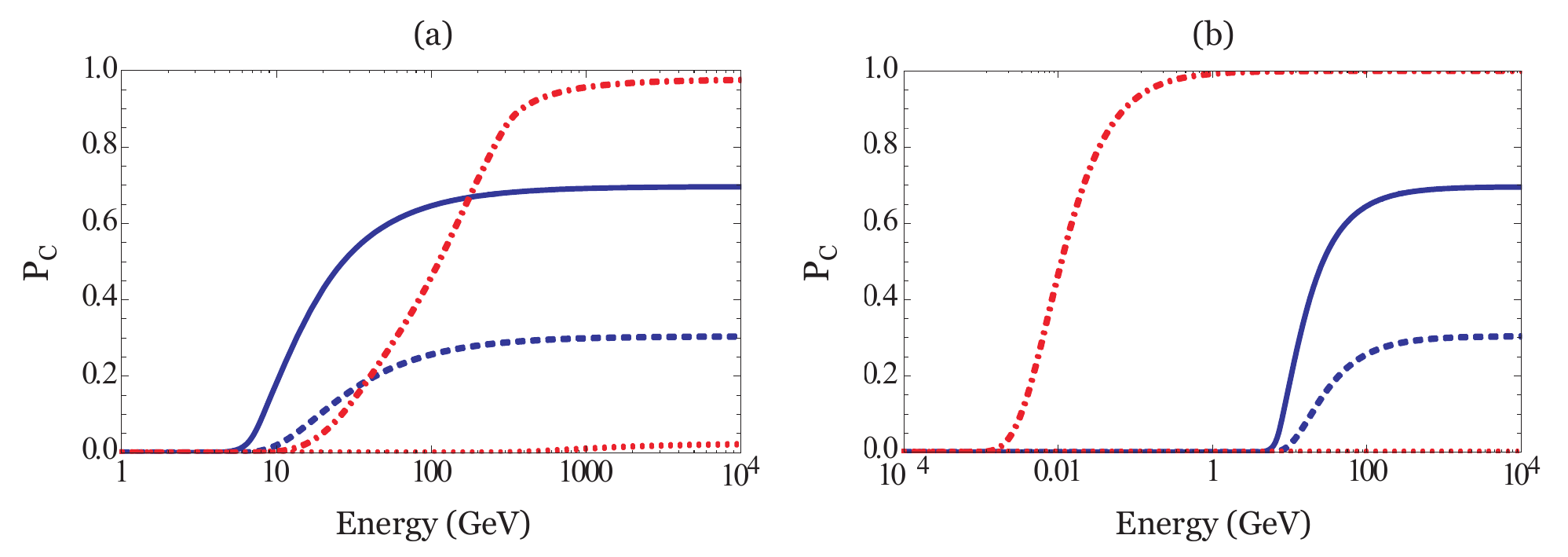}
\caption{The level-crossing probabilities $P_C$; our chosen value of
  $\theta_{13} = 8.8^{\circ}$ is used in fig (a), while fig (b) shows
  the corresponding result for $\theta_{13} = 0.1^{\circ}$. The first
  ($\theta_{13}$-driven) level crossing probabilities ($P_C^h$) are
  shown by the red dot-dashed (neutrino) and dotted (antineutrino)
  lines, while the second ($\theta_{12}$-driven) crossing
  probabilities ($P_C^l$) are shown by the blue bold (neutrino) and
  dashed (antineutrino) lines.}
\label{fig:Pc}
\end{figure}

The level crossing probabilities, for neutrinos and antineutrinos, are
plotted as a function of neutrino energy in
figure~\ref{fig:Pc}. Figure~\ref{fig:Pc}(a) shows the probabilities
used here, assuming $\theta_{13} = 8.8^{\circ}$, as suggested by
recent T2K~\cite{abe2011indication} and the Daya
Bay~\cite{an2012observation} results, while figure~\ref{fig:Pc}(b)
shows the probabilities if we instead allow $\theta_{13} =
0.1^{\circ}$.  Note that for very small $\theta_{13}$, $P_C^h$ for
neutrinos approaches 1 for all energies (i.e. the resonance is
completely non-adiabatic).  This is to be expected because as
$\theta_{13} \rightarrow 0$ the heaviest vacuum eigenstate becomes
completely decoupled, and does not mix with the other mass
eigenstates.  For antineutrinos $P_C^h\simeq 0$, as the resonance
condition is never met.  We also see that the lower crossing
probabilities, $P_C^l$ are comparatively larger for neutrinos than
antineutrinos, as $\bar{\nu}_e$ is produced in the lightest matter
mass eigenstate and, again, does not undergo a resonance.

\begin{figure}[t]
\vspace*{-0.4cm}
\hspace{-0.8cm}
\includegraphics[scale=0.8]{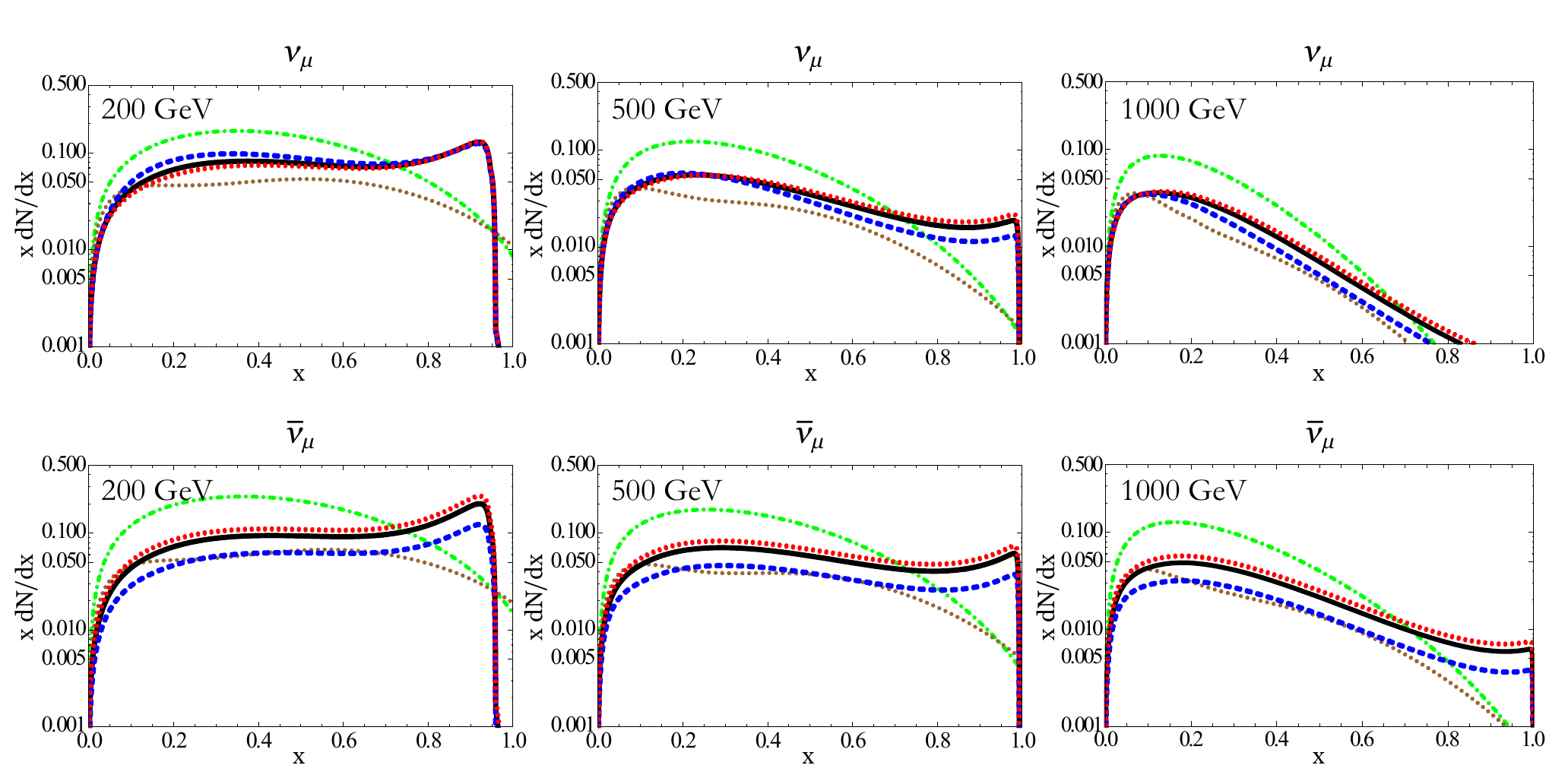}
\caption{The final muon-flavour (anti)neutrino energy spectra, per
  annihilation, after propagation from the Sun's centre to Earth, for
  $M_\chi=200$, 500, 1000 GeV.  The curves denote different
  annihilation branching ratios: equal branching to all lepton
  flavours (black, bold), branching only to electron flavour (blue,
  dashed), branching evenly to $\mu$ and $\tau$ flavours (red, bold
  dotted). Also shown for comparison are the spectra arising from DM
  annihilation directly to $\tau^+ \tau^-$ (green, dot-dashed) and to
  $W^+ W^-$ (brown, dotted).  We set $\theta_{13} \simeq 8.8^{\circ}$, and $\mu =$ 1.2.}
\label{fig:AfterEvol}
\end{figure}

We apply eq. (\ref{eq:totalPc}) to obtain the neutrino fluxes at the
edge of the Sun, in the vacuum mass basis.  The neutrinos propagate
from the Sun to the Earth in these vacuum states, and can then be
converted back to the flavour basis.  The final spectra are
shown in figure~\ref{fig:AfterEvol}.

\subsection{Final spectra}

The final neutrino spectra, per annihilation, for the electroweak
bremsstrahlung annihilation mode are shown in
figure~\ref{fig:AfterEvol}, for various choices of the branching
ratios to the lepton flavours. Also shown, for comparison, are the
spectra that would result from DM annihiation to $W^+ W^-$ and $\tau^+
\tau^-$ final states.  These comparison spectra were calculated using
the same procedure described in section 2.

We note that while the initial spectra at production were sensitive to
flavour branching ratios chosen, neutrino flavour mixing in the Sun
considerably reduces these differences.  For example, for branching
only to the electron flavour, the only $\nu_\mu$ and $\nu_\tau$
components at production are secondary neutrinos.  However, the flavour
mixing that takes places via the level crossings serves to even out
the distribution across the three flavours. The final spectra have a
similar shape, irrespective of the branching ratios chosen, with
magnitudes that vary by only a factor of $\lesssim 2$.

Scattering and absorption have a strong effect on the spectra for
$M_\chi = 1000$~GeV, but very little effect for $M_\chi = 200$~GeV.
This is expected as absorption depends exponentially on the neutrino
energy.  For large $M_\chi$, the high energy regions of the spectra
are strongly damped, removing the high energy peak.  This is a bigger
effect for neutrinos than for antineutrinos due to the different
values of $\mathcal{E}$ --- the scattering in the Sun affects
neutrinos more strongly than their antiparticles.  Overall, absorption
significantly reduces the number of detectable neutrinos produced per
annihilation.  For the energies we consider, regeneration, which
reintroduces neutrinos creating a pile-up at lower energies, is a
smaller effect.

\begin{figure}[t]
\begin{center} 
\vspace*{-1cm}
\hspace{-1.5cm}
\includegraphics[scale=0.6]{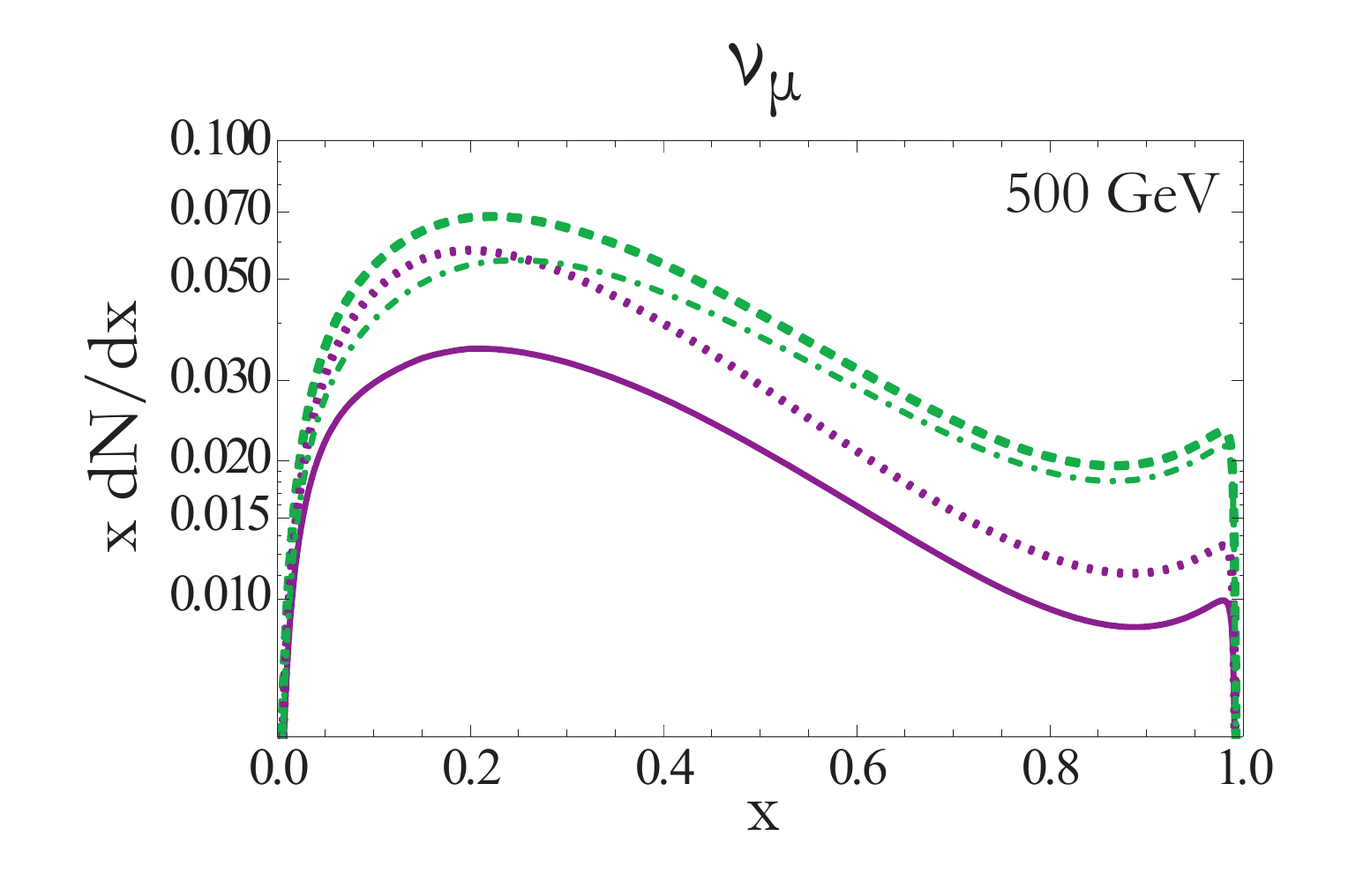}
\caption{The $\nu_{\mu}$ spectra for $M_{\chi} = 500$ GeV, for
  different choices of of $\theta_{13}$, and where $\mu =$ 1.2.  For branching only to
  $e$-flavoured leptons the spectra are shown for $\theta_{13} =
  0.1^{\circ}$ (purple, solid) and $\theta_{13} = 8.8^{\circ}$
  (purple, dotted).  For branching to $\mu$ and $\tau$ flavours
  equally the curves correspond to $\theta_{13} = 0.1^{\circ}$ (green,
  dashed) and $\theta_{13} = 8.8^{\circ}$ (green, dot-dashed). Note
  that there is no significant dependence on $\theta_{13}$ for the
  case of equal branching to all flavours.  }
\label{fig:theta13plots}
\end{center}
\end{figure}

These plots in figure~\ref{fig:AfterEvol} were produced assuming a
value for $\theta_{13}$ of $8.8^{\circ}$.  For the purposes of
comparison, we also computed spectra for $\theta_{13} = 0.1^{\circ}$.
In the $\theta_{13}\simeq 0$ limit, the evolution through the high
energy $\theta_{13}$-driven resonance is completely non-adiabatic,
with $P_C^h \simeq 1$ for all energies of interest, while for the
larger mixing angle the resonance is (partially) adiabatic for
energies $\lesssim$ 1000 GeV.  In figure~\ref{fig:theta13plots}, we
compare the neutrino spectra for the two choices of $\theta_{13}$.
Decreasing $\theta_{13}$ decreases the adiabaticity of the resonance,
and increases the differences between the spectra for each flavour.
This effect is most significant at low energies, where the
level-crossing probabilities are small (see figure~\ref{fig:Pc}),
while at high energies $P_C^h \rightarrow 1$ irrespective of the value
of $\theta_{13}$.  For antineutrinos, there is very little sensitivity
to $\theta_{13}$ as the $\theta_{13}$-resonance condition is never
met.
From here on, we shall assume equal branching to all three flavours
($f_e=f_\mu=f_\tau$), which renders the results insensitive to the
value of $\theta_{13}$.

\section{Signals at Earth}


We now compare the signal from EW bremsstrahlung annihilation, to that
arising in models in which DM annihilates instead to 2-body final
states such as $W^+W^-$ or $\tau^+\tau^-$.  These channels have been
chosen for comparison as they both have relatively hard neutrino
spectra, with a significant flux at large values of $x$.  As a result
the existing limits on these
channels~\cite{abe2011indirect,abbasi2010limits,abbasi2009limits} are
more stringent than those with softer neutrino spectra.  As seen in
figure~\ref{fig:AfterEvol}, the EW bremsstrahlung, $ W^+ W^-$ and
$\tau^+ \tau^-$ channels all have broad energy spectra, of similar
shape and scale.  However the EW bremsstrahlung channel has a
significant high energy peak at $x \sim 1$, which will improve its
detectability compared to the $W^+ W^-$ mode for which limits are
commonly determined.  We estimate relative event rates in the following
subsection.

The spectra in figure~\ref{fig:AfterEvol} have been computed per
annihlation; the absolute normalisation of the flux is specified by
eq.~\ref{eq:diffflux}.  (If we assume capture-annihilation equilibrium,
the absolute normalization of the event rate is set by the capture
rate alone times the annihilation branching ratio of the channel of
interest.)  However, for a given annihilation rate (which we leave as
a free parameter) it is useful to compare the \emph{ratio} of event
rates for possible annihilation channels.  We can thus assess the
detectability of the EW bremsstrahlung annihilation channel relative
to that for the comparison 2-body final state channels, $W^+W^-$ and
$\tau^+\tau^-$, by comparing their relative event rates.


\subsection{Muon event rates}

We consider muon event rates in a detector such as IceCube.  Although
IceCube is able to detect neutrinos of all flavours via cascade
events, we consider here only the muon-track events as they have the
greatest detection probability, since the long muon range in ice
allows the detection of muons created well outside the detector.

We estimate event rates following a similar procedure to that in
refs.~\cite{gandhi1996ultrahigh,dutta2001propagation,beacom2003measuring,bell2011enhanced}.
A muon neutrino can undergo charged current scattering with a nucleon
within the ice of the detector, to produce a muon with energy $E_{\mu}
= E_{\nu} (1-y)$, where $y$ is the charged current inelasticity
parameter.  This muon will then propagate through the ice and lose
energy, with a range given by
\begin{equation}
\label{eq:Rmu}
R_{\mu}(E_{\mu}, E_{\mu}^{\mathrm{thr}}) = \frac{1}{\beta} \, \mathrm{ln} \, \bigg [\frac{\alpha + \beta E_{\mu}}{\alpha + \beta E_{\mu}^{\mathrm{thr}}} \bigg] \, \, ,
\end{equation}
where $\alpha = 2.0 \, \mathrm{MeV} \, \mathrm{cm}^2 /\mathrm{g}$, $\beta = 4.2 \times 10^{-6} \mathrm{cm}^2/\mathrm{g}$, and $E_{\mu}^{\mathrm{thr}}$ is the muon energy detection threshold, taken to be 100 GeV. Therefore, the probability that a neutrino with energy $E_{\nu}$ creates a muon that is detected with an energy above $E_{\mu}^{\mathrm{thr}}$ is given by
\begin{equation}
\label{eq:muonProb}
P(E_{\nu}, E_{\mu}^{\mathrm{thr}}) = \rho_N N_A \sigma_{\mathrm{nucleon}}(E_{\nu}) R_{\mu}(E_{\nu} (1-y), E_{\mu}^{\mathrm{thr}}) \, \, ,
\end{equation}
where $\rho_N$ is the target nucleon density, $N_A$ is Avogadro's
number, and $\sigma_{\mathrm{nucleon}}(E_{\nu})$ is the
charged current scattering cross section in ice. Finally, the muon
track event rate is proportional to
\begin{equation}
\label{eq:muonrate}
\int \frac{dN(E_{\nu})}{dE_{\nu}} P(E_{\nu},E_{\mu}^{\mathrm{thr}}) A^{\mathrm{eff}} dE_{\nu},
\end{equation}
where $A^{\mathrm{eff}}$ is the effective area of the detector; for
IceCube this is $\sim 1 \mathrm{km}^2$ \cite{ahrens2004sensitivity}.
The scattering cross section $\sigma_{\mathrm{nucleon}}(E_{\nu})$ is
proportional to $E_{\nu}$ and, in ice, the neutrino cross section is
approximately twice as large as that for antineutrinos. We choose the
average values of $y$ as tabulated in ref. \cite{gandhi1996ultrahigh},
which are approximately $\sim$0.45 and $\sim$0.35 for neutrinos and
antineutrinos respectively.  As a result, the lower bound on the
(anti)neutrino energies that will produce observable muons is
\begin{equation}
\begin{aligned}
E_{\nu_{\mu}} & \geq E_{\mu}^{\mathrm{thr}}/(1-y_{\nu}), \hspace{1.5cm} & 
E_{\bar{\nu}_{\mu}} & \geq E_{\mu}^{\mathrm{thr}}/(1-y_{\bar{\nu}}), \\
& \simeq 182 \mathrm{GeV}, & & \simeq 154 \mathrm{GeV} \,\, . \\
\end{aligned}
\end{equation}

In table~\ref{tab:muonevents}, we show the ratio of muon-track events
for various annihilation modes, assuming the same overall
normalisation.  For example, by setting the branching ratio in the
table equal to 1, we can compare the event rate for a model in which
the annihilation proceeds via bremsstrahlung with that for a model in
which the annihilation is solely to $W^+W^-$ or $\tau^+\tau^-$,
assuming the same total annihilation rate. 

We see that, for a
given annihilation rate, annihilation via the EW bremsstrahlung
channel would produce more events than annihilation via a $W^+ W^-$
channel, for all masses considered, with a ratio of $\sim 4$ at
$M_{\chi}=200$ GeV.  The ratio is largest for low dark matter masses,
because only the high energy regions of the spectra are above the
detection threshold, and hence the high energy peak in the EW
bremsstrahlung spectrum dominates.  The ratio is smaller for large
$M_\chi$, as the detection threshold then corresponds to smaller
values of $x$, and absorption effects damp the spectra at high energy.
There is only a mild dependence of these results on the value of the
mass degeneracy, $\mu = M_{\eta}^2/M_{\chi}^2$.  For larger values of
$\mu$ the high energy peak is diminished, but more mid-to-high energy
neutrinos are produced, such that the total number of detectable events
remains approximately constant.

Detection prospects for annihilation channels which produce hard
spectra, such as $W^+ W^-$, are much better than those for annihilation
channels which produce much softer neutrino spectra, such as
$\bar{q}q$.  Given that annihilation via the bremsstrahlung channel
would result in more events than annihilation to $W^+ W^-$, its
detection prospects appear promising.  The experimental sensitivity
for a given annihilation channel is usually expressed in terms of a limit
on the DM-nucleon scattering cross section which controls the absolute
event rate (assuming capture-annihilation equilibrium).  For
annihilation to purely $W^+W^-$, the projected IceCube sensitivity (180 days of
the full 86 string configuration) to spin-dependent DM-nucelon
scattering is $\sigma_{\chi N}^{SD} \sim \mathrm{few} \times 10^{-5}
~\mathrm{pb}$~\cite{abbasi2012multiyear} for the DM mass range we
consider.  Since annihilation purely via the bremmstrahlung channel
will produce more events than the $W^+W^-$ channel by a factor of a
few, we estimate a sensitivity of $\sigma_{\chi N}^{SD} \sim 1 \times
10^{-5}~\mathrm{ pb}$ for this channel (see also ref.~\cite{fukushima2012brem}).

\begin{table}
\newcommand\T{\rule{0pt}{3.4ex}}
\newcommand\B{\rule[-2.0ex]{0pt}{0pt}}
\begin{center}
\label{tab:muonevents}
	\begin{tabular}{| c || c | c || c | c || c | c ||}
		\hline
 		  & \multicolumn{2}{c||}{200 GeV} & \multicolumn{2}{c||}{500 GeV} & \multicolumn{2}{c||}{1000 GeV}  \\ 
  & $\mu=1.2$ & $\mu=5$   & $\mu=1.2$ & $\mu=5$ & $\mu=1.2$ & $\mu=5$ \\ \hline
  \T \B	$\frac{\Phi_\nu(\mathrm{brem})}{\Phi_\nu (\tau^+ \tau^-)} 
                  /\frac{\mathrm{Br(brem)}}{\mathrm{Br}(\tau^+ \tau^-)}$ & 1.8 & 2.1 & 1.2 & 1.2 & 0.6 & 0.7 \\  \hline 
    \T \B 		$\frac{\Phi_\nu(\mathrm{brem})}{\Phi_\nu (W^+ W^-)} 
                  /\frac{\mathrm{Br(brem)}}{\mathrm{Br}(W^+W^-)}$ & 3.8 & 4.6 & 2.4 & 2.6 & 1.9 & 2.0 \\ \hline
	\end{tabular}
	\caption{The ratio of $\mu^{\pm}$ muon-track events in an
          IceCube-like detector, for various annihilation channels.
          We have assumed the same overall normalization for each
          possible annihilation mode, that is, assuming
          capture-annihilation equilibrium to hold, we assume the same
          capture rate for each case. For the EW bremsstrahlung
          annihilation channel, equal branching ratios to all lepton
          flavours was assumed ($f_e=f_\mu=f_\tau$).}
\end{center}
\end{table}

\subsection{Parameter constraints}
\label{parameters}


We have presented spectra per annihilation, in the spirit of
ref.~\cite{cirelli2005spectra}, choosing to leave the absolute
normalization as a free parameter.  However, the detection prospects
for a given DM model depend on the absolute event rate, which is
determined by eq.~\ref{eq:diffflux} and eq.~\ref{eq:annrate} and, in
general, will depend on both the DM-nucleon scattering and the DM
annihilation cross sections.  However, importantly, if
capture-annihilation equilibrium is established, then the event rate
is controlled by the DM-nucleon scattering cross section alone (modulo
astrophysical uncertainties in the local dark matter density and
velocity).  We now discuss whether this condition is expected to
hold in the DM models of interest for electroweak bremsstrahlung.

Expressions for the capture and annihilation rate are given, e.g., in
ref.~\cite{bertone2005particle}.  In order for equilibrium to be
established, the product of (i) the DM-nucleon scattering and (ii) the
DM annihilation cross section must be sufficiently large.  If one
assumes a spin dependent DM-nucleon scattering cross section of $\sim
10^{-5}~\textrm{pb}$ (the IceCube sensitivity estimated above), an
annihilation cross section of $\langle \sigma v \rangle \gtrsim
10^{-2} \langle \sigma v \rangle_{\textrm{th}}$ is required for
capture-annihilation equilibrium to be reached within the lifetime of
the solar system, where $\langle \sigma v \rangle_{\textrm{th}} \simeq
2 \times 10^{-26} \textrm{cm}^3/\textrm{s}$ is the annihilation cross
section expected for a thermal relic \cite{steigman2012precise}.  We discuss each of these cross
sections in turn.

The bremsstrahlung annihilation cross section is expected to be
approximately two orders of magnitude smaller than that for a thermal
relic, $\langle \sigma v \rangle_\textrm{brem} \sim 10^{-2} \langle
\sigma v \rangle_{\textrm{th}}$.~\footnote{Larger annihilation cross
  sections would imply either a subdominant DM component, or that the
  DM was produced not by thermal freezeout but by some more exotic
  mechanism.  A number of recent papers have considered indirect
  detection and set upper limits on the EW bremsstrahlung annihilation
  cross section based on observations of antiprotons and gamma rays.
  For instance refs.~\cite{garny2012dark,bell2011dark} both find
  limits of approximately $\langle \sigma v \rangle \lesssim 10^{-24}
  - 10^{-25} \textrm{cm}^3/\textrm{s}$ over the mass range 100-300
  GeV.  Ref.~\cite{Bringmann:2012vr} performs a search for a
  bremsstrahlung annihilation signal in the Fermi LAT gamma ray data
  and sets bounds of approximately $\langle \sigma v \rangle \lesssim
  10^{-27} \textrm{cm}^3/\textrm{s}$ over the same mass range.} This is because the annihilation
cross section at thermal freezeout is dominated not by the
bremsstrahlung channels, but instead by annihilation to 2-body final
states.  The 3-body bremsstrahlung processes dominate the
annihilation in the late universe, because the 2-body final states are
p-wave suppressed by a factor $v^2 \sim 10^{-6}$; in the early
universe $v$ is much larger and the 2-body final state is not strongly
suppressed.  See the discussion in section 4 of
ref~\cite{Bringmann:2012vr} for further details.

For a purely leptophilic model, the DM couplings to quarks arise at
loop level.  For the coupling of eq.~\ref{eq:Caomodel} the DM-nucleon
interaction arises at 1-loop level through a coupling to the Z boson,
resulting in an effective interaction (\cite{cao2009dark})
\begin{equation}
\mathcal{L}_A = \frac{\mathcal{G}}{M_Z^2}(\bar{\chi} \gamma^{\mu} \gamma_5 \chi)(\bar{q} \gamma_{\mu} \gamma_5 q),
\end{equation}
where $\mathcal{G}$ is the loop-induced form factor.  The resulting
spin-dependent DM-nucleon cross section is shown in figure 2(a) of
ref.\cite{cao2009dark}, and can take values $\lesssim 10^{-6}~\mathrm{pb}$.


Therefore, for the purely leptophilic model described by the coupling
of eq.~\ref{eq:Caomodel}, the product of the DM-nucleon scattering and
DM annihilation cross sections are sufficiently small that
capture-annihilation equilibrium would not hold (although we would be
close to this limit, with event rates suppressed by a factor of less
than 10 for some parameters).
However, if the model were not purely leptophilic, the DM-nucleon
cross section could easily be larger, such that capture-annihilation
equilibrium would hold.  For instance, couplings identical to that we
consider arise in SUSY models.  Embedding the toy model of
eq.~\ref{eq:Caomodel} into a full theory (e.g. the MSSM, or other
theory beyond the standard model) is a more realistic scenario.  In
that case it is likely that several annihilation channels would be
open, which would need to be combined with model dependent branching
ratios.  In this paper we take a purely phenomenological approach to
study the EW bremsstrahlung signal in isolation.



\section{Conclusions}

High-energy neutrinos produced by the annihilation of dark matter captured in
the Sun offer a useful dark matter search strategy.  By comparing the
predicted neutrino fluxes with future detections, we can potentially
distinguish between different DM annihilation channels, thus
providing information to discriminate among competing DM models.

DM annihilation is usually assumed to be dominated by annihilation to
2-body final states.  However, there are important DM annihilation
processes with 3-body final states, namely bremsstrahlung processes in
which a $W$, $Z$ or $\gamma$ is radiated.  The electroweak
bremsstrahlung processes $\chi \chi \rightarrow \bar{f}f Z$ and $\chi
\chi \rightarrow f \nu W$ have been shown to be the dominant DM
annihilation mode in certain popular models in which the lowest order
annihilation processes are helicity suppressed.  In the context of
solar WIMP annihlation, these electroweak bremsstrahlung annihilation
channels produce large, energetic, fluxes of neutrinos.

We have computed the neutrino energy spectra resulting from DM
annihilation via the electroweak bremsstrahlung channel.  This
consists of primary neutrinos produced directly in the annihilations,
which have a distinguishing peak near the endpoint $E_\nu = M_\chi$,
and secondary neutrinos produced in decays of annihilation products.
By comparing the event rates for the electroweak bremsstrahlung
annihilation channel with those for annihilation instead to $W^+W^-$,
we have shown that the EW bremsstrahlung channel has the larger event
rate per annihilation and hence promising detection prospects.

\acknowledgments

AJB was supported by the Commonwealth of Australia and NFB by the Australian Research Council. We thank Thomas Weiler, Kalliopi Petraki, James Dent, Lawrence Krauss, Nicholas Setzer and Martin White for helpful discussions.

\bibliography{Bib}

\providecommand{\href}[2]{#2}\begingroup\raggedright\begin{thebibliography}{10}

\bibitem{silk1985photino}
J.~Silk, K.~Olive, and M.~Srednicki, {\it {The photino, the sun, and
  high-energy neutrinos}},  {\em Phys.Rev.Lett.} {\bf 55} (1985) 257--259.

\bibitem{press1985capture}
W.~H. Press and D.~N. Spergel, {\it {Capture by the Sun of a Galactic
  Population of Weakly Interacting Massive Particles}},  {\em Astrophys.J.}
  {\bf 296} (1985) 679--684.

\bibitem{freese1986can}
K.~Freese, {\it {Can scalar neutrinos or massive Dirac neutrinos be the missing
  mass?}},  {\em Phys.Lett.} {\bf B167} (1986) 295--300.

\bibitem{krauss1986solar}
L.~Krauss, M.~Srednicki, and F.~Wilczek, {\it Solar system constraints and
  signatures for dark-matter candidates},  {\em Phys.Rev.} {\bf D33} (1986)
  2079.

\bibitem{gaisser1986limits}
T.~Gaisser, G.~Steigman, and S.~Tilav, {\it {Limits on cold-dark-matter
  candidates from deep underground detectors}},  {\em Phys.Rev.} {\bf D34}
  (1986) 2206.

\bibitem{gould1987resonant}
A.~Gould, {\it {Resonant Enhancements in WIMP Capture by the Earth}},  {\em
  Astrophys.J.} {\bf 321} (1987) 571.

\bibitem{srednicki1987high}
M.~Srednicki, K.~Olive, and J.~Silk, {\it {High-energy neutrinos from the sun
  and cold dark matter}},  {\em Nucl.Phys.} {\bf B279} (1987) 804--823.

\bibitem{kamionkowski1991energetic}
M.~Kamionkowski, {\it Energetic neutrinos from heavy-neutralino annihilation in
  the sun},  {\em Phys.Rev.} {\bf D44} (1991) 3021--3042.

\bibitem{cirelli2005spectra}
M.~Cirelli, N.~Fornengo, T.~Montaruli, I.~A. Sokalski, A.~Strumia, et~al., {\it
  {Spectra of Neutrinos from Dark Matter Annihilations}},  {\em Nucl.Phys.}
  {\bf B727} (2005) 99--138,
  [\href{http://xxx.lanl.gov/abs/hep-ph/0506298}{{\tt hep-ph/0506298}}].

\bibitem{blennow2008neutrinos}
M.~Blennow, J.~Edsj{\"o}, and T.~Ohlsson, {\it {Neutrinos from WIMP
  annihilations obtained using a full three-flavor Monte Carlo approach}},
  {\em JCAP} {\bf 01} (2008) 021,
  [\href{http://xxx.lanl.gov/abs/0709.3898}{{\tt arXiv:0709.3898}}].

\bibitem{barger2002indirect}
V.~Barger, F.~Halzen, D.~Hooper, and C.~Kao, {\it {Indirect search for
  neutralino dark matter with high energy neutrinos}},  {\em Phys.Rev.} {\bf
  D65} (2002) 075022, [\href{http://xxx.lanl.gov/abs/hep-ph/0105182v1}{{\tt
  hep-ph/0105182v1}}].

\bibitem{Lehnert:2010vb}
R.~Lehnert and T.~J. Weiler, {\it {Flavor Sensitivity to $\theta_{13}$ and the
  Mass Hierarchy for neutrinos from Solar WIMP Annihilation}},
  \href{http://xxx.lanl.gov/abs/1002.2441}{{\tt arXiv:1002.2441}}.

\bibitem{erkoca2009muon}
A.~Erkoca, M.~Reno, and I.~Sarcevic, {\it {Muon fluxes from dark matter
  annihilation}},  {\em Phys.Rev.} {\bf D80} (2009) 043514,
  [\href{http://xxx.lanl.gov/abs/0906.4364}{{\tt arXiv:0906.4364}}].

\bibitem{belotsky2009muon}
K.~Belotsky, M.~Khlopov, and C.~Kouvaris, {\it {Muon flux limits for Majorana
  dark matter from strong coupling theories}},  {\em Phys. Rev.} {\bf D79}
  (2009) 083520, [\href{http://xxx.lanl.gov/abs/0810.2022}{{\tt
  arXiv:0810.2022}}].

\bibitem{barger2010fermion}
V.~Barger, J.~Kumar, D.~Marfatia, and E.~M. Sessolo, {\it {Fermion WIMPless
  Dark Matter at DeepCore and IceCube}},  {\em Phys.Rev.} {\bf D81} (2010)
  115010, [\href{http://xxx.lanl.gov/abs/1004.4573}{{\tt arXiv:1004.4573}}].

\bibitem{abe2011indirect}
{\bf Super-Kamiokande} Collaboration, K.~Abe et~al., {\it {An Indirect Search
  for WIMPs in the Sun using 3109.6 days of upward-going muons in
  Super-Kamiokande}},  \href{http://xxx.lanl.gov/abs/astro-ph/1108.3384}{{\tt
  astro-ph/1108.3384}}.

\bibitem{abbasi2010limits}
{\bf IceCube} Collaboration, R.~Abbasi et~al., {\it {Limits on a muon flux from
  Kaluza-Klein dark matter annihilations in the Sun from the IceCube 22-string
  detector}},  {\em Phys.Rev.} {\bf D81} (2010) 057101,
  [\href{http://xxx.lanl.gov/abs/astro-ph/0910.4480v1}{{\tt
  astro-ph/0910.4480v1}}].

\bibitem{abbasi2009limits}
{\bf IceCube} Collaboration, R.~Abbasi et~al., {\it {Limits on a muon flux from
  neutralino annihilations in the Sun with the IceCube 22-string detector}},
  {\em Phys.Rev.Lett.} {\bf 102} (2009) 201302,
  [\href{http://xxx.lanl.gov/abs/astro-ph/0902.2460v3}{{\tt
  astro-ph/0902.2460v3}}].

\bibitem{deyoung2011particle}
{\bf IceCube} Collaboration, T.~DeYoung, {\it {Particle physics in ice with
  IceCube DeepCore}},  {\em Nucl.Instrum.Meth. A} (2011)
  [\href{http://xxx.lanl.gov/abs/1112.1053}{{\tt arXiv:1112.1053}}].

\bibitem{bell2011wrev}
N.~Bell, J.~Dent, A.~Galea, T.~Jacques, L.~Krauss, and T.~Weiler, {\it {W/Z
  Bremsstrahlung as the Dominant Annihilation Channel for Dark Matter,
  Revisited}},  {\em Phys.Lett.} {\bf B706} (2011)
  [\href{http://xxx.lanl.gov/abs/1104.3823}{{\tt arXiv:1104.3823}}].

\bibitem{bell2011w}
N.~F. Bell, J.~B. Dent, T.~D. Jacques, and T.~J. Weiler, {\it {W/Z
  Bremsstrahlung as the Dominant Annihilation Channel for Dark Matter}},  {\em
  Phys.Rev.} {\bf D83} (2011) 013001,
  [\href{http://xxx.lanl.gov/abs/1009.2584}{{\tt arXiv:1009.2584}}].

\bibitem{bergstrom1989radiative}
L.~Bergstr{\"o}m, {\it {Radiative Processes in Dark Matter Photino
  Annihilation}},  {\em Phys.Lett.} {\bf B225} (1989) 372.

\bibitem{flores1989radiative}
R.~Flores, K.~Olive, and S.~Rudaz, {\it {Radiative processes in LSP
  annihilation}},  {\em Phys.Lett.} {\bf B232} (1989) 377--382.

\bibitem{baltz2003detection}
E.~Baltz and L.~Bergstr{\"o}m, {\it {Detection of Leptonic Dark Matter}},  {\em
  Phys.Rev.} {\bf D67} (2003) 043516,
  [\href{http://xxx.lanl.gov/abs/hep-ph/0211325}{{\tt hep-ph/0211325}}].

\bibitem{bringmann2008new}
T.~Bringmann, L.~Bergstr{\"o}m, and J.~Edsj{\"o}, {\it {New Gamma-Ray
  Contributions to Supersymmetric Dark Matter Annihilation}},  {\em JHEP} {\bf
  0801} (2008) 049, [\href{http://xxx.lanl.gov/abs/hep-ph/0710.3169v3}{{\tt
  hep-ph/0710.3169v3}}].

\bibitem{bergstrom2008new}
L.~Bergstr{\"o}m, T.~Bringmann, and J.~Edsj{\"o}, {\it {New Positron Spectral
  Features from Supersymmetric Dark Matter: A Way to Explain the PAMELA
  Data?}},  {\em Phys.Rev.} {\bf D78} (2008) 103520,
  [\href{http://xxx.lanl.gov/abs/astro-ph/0808.3725v3}{{\tt
  astro-ph/0808.3725v3}}].

\bibitem{barger2009generic}
V.~Barger, Y.~Gao, W.~Keung, and D.~Marfatia, {\it Generic dark matter
  signature for gamma-ray telescopes},  {\em Phys. Rev. D} {\bf 80} (Sep, 2009)
  063537, [\href{http://xxx.lanl.gov/abs/hep-ph/0906.3009v2}{{\tt
  hep-ph/0906.3009v2}}].

\bibitem{goldberg1983constraint}
H.~Goldberg, {\it Constraint on the photino mass from cosmology},  {\em
  Phys.Rev.Lett.} {\bf 50} (1983) 1419--1422.

\bibitem{krauss1983new}
L.~Krauss, {\it {New constraints on ``INO" masses from cosmology}},  {\em
  Nucl.Phys.} {\bf B227} (1983) 556--569.

\bibitem{bell2011dark}
N.~F. Bell, J.~B. Dent, T.~D. Jacques, and T.~J. Weiler, {\it {Dark Matter
  Annihilation Signatures from Electroweak Bremsstrahlung}},  {\em Phys.Rev.}
  {\bf D84} (2011) 103517, [\href{http://xxx.lanl.gov/abs/1101.3357}{{\tt
  arXiv:1101.3357}}].

\bibitem{bell2008electroweak}
N.~F. Bell, J.~B. Dent, T.~D. Jacques, and T.~J. Weiler, {\it {Electroweak
  Bremsstrahlung in Dark Matter Annihilation}},  {\em Phys.Rev.} {\bf D78}
  (2008) 083540, [\href{http://xxx.lanl.gov/abs/0805.3423}{{\tt
  arXiv:0805.3423}}].

\bibitem{barger2012bremsstrahlung}
V.~Barger, W.~Keung, and D.~Marfatia, {\it {Bremsstrahlung in dark matter
  annihilation}},  {\em Phys.Lett.} {\bf B707} (2012)
  [\href{http://xxx.lanl.gov/abs/1111.4523}{{\tt arXiv:1111.4523}}].

\bibitem{ciafaloni1999sudakov}
P.~Ciafaloni and D.~Comelli, {\it {Sudakov Effects in Electroweak
  Corrections}},  {\em Phys.Lett.} {\bf B446} (1999) 278--284,
  [\href{http://xxx.lanl.gov/abs/hep-ph/9809321}{{\tt hep-ph/9809321}}].

\bibitem{ciafaloni2010tev}
P.~Ciafaloni and A.~Urbano, {\it {TeV scale Dark Matter and Electroweak
  Radiative Corrections}},  {\em Phys.Rev.} {\bf D82} (2010) 043512,
  [\href{http://xxx.lanl.gov/abs/1001.3950}{{\tt arXiv:1001.3950}}].

\bibitem{ciafaloni2011weak}
P.~Ciafaloni, D.~Comelli, A.~Riotto, F.~Sala, A.~Strumia, and A.~Urbano, {\it
  {Weak corrections are relevant for dark matter indirect detection}},  {\em
  JCAP} {\bf 03} (2011) 019, [\href{http://xxx.lanl.gov/abs/1009.0224}{{\tt
  arXiv:1009.0224}}].

\bibitem{ciafaloni2011importance}
P.~Ciafaloni, M.~Cirelli, D.~Comelli, A.~De~Simone, A.~Riotto, and A.~Urbano,
  {\it {On the importance of electroweak corrections for Majorana dark matter
  indirect detection}},  {\em JCAP} {\bf 06} (2011) 018,
  [\href{http://xxx.lanl.gov/abs/1104.2996}{{\tt arXiv:1104.2996}}].

\bibitem{ciafaloni2011initial}
P.~Ciafaloni, M.~Cirelli, D.~Comelli, A.~De~Simone, A.~Riotto, and A.~Urbano,
  {\it {Initial state radiation in Majorana Dark Matter annihilations}},  {\em
  JCAP} (2011) 034, [\href{http://xxx.lanl.gov/abs/1107.4453}{{\tt
  arXiv:1107.4453}}].

\bibitem{garny2011antiproton}
M.~Garny, A.~Ibarra, and S.~Vogl, {\it {Antiproton constraints on dark matter
  annihilations from internal electroweak bremsstrahlung}},  {\em JCAP} {\bf
  07} (2011) 028, [\href{http://xxx.lanl.gov/abs/1105.5367}{{\tt
  arXiv:1105.5367}}].

\bibitem{garny2012dark}
M.~Garny, A.~Ibarra, and S.~Vogl, {\it {Dark matter annihilations into two
  light fermions and one gauge boson: general analysis and antiproton
  constraints}},  \href{http://xxx.lanl.gov/abs/1112.5155}{{\tt
  arXiv:1112.5155}}.

\bibitem{ciafaloni2012electroweak}
P.~Ciafaloni, D.~Comelli, A.~De~Simone, A.~Riotto, and A.~Urbano, {\it
  {Electroweak bremsstrahlung for wino-like Dark Matter annihilations}},
  \href{http://xxx.lanl.gov/abs/1202.0692}{{\tt arXiv:1202.0692}}.

\bibitem{kachelriess2009role}
M.~Kachelriess, P.~Serpico, and M.~Solberg, {\it {Role of electroweak
  bremsstrahlung for indirect dark matter signatures}},  {\em Phys. Rev.} {\bf
  D80} (2009) 123533, [\href{http://xxx.lanl.gov/abs/0911.0001}{{\tt
  arXiv:0911.0001}}].

\bibitem{Bringmann:2012vr}
T.~Bringmann, X.~Huang, A.~Ibarra, S.~Vogl, and C.~Weniger, {\it {Fermi LAT
  Search for Internal Bremsstrahlung Signatures from Dark Matter
  Annihilation}},  \href{http://xxx.lanl.gov/abs/1203.1312}{{\tt
  arXiv:1203.1312}}.

\bibitem{griest1987cosmic}
K.~Griest and D.~Seckel, {\it {Cosmic Asymmetry, Neutrinos and the Sun}},  {\em
  Nucl.Phys.} {\bf B283} (1987) 681.

\bibitem{bottino2002does}
A.~Bottino, G.~Fiorentini, N.~Fornengo, B.~Ricci, S.~Scopel, and F.~Villante,
  {\it {Does solar physics provide constraints to weakly interacting massive
  particles?}},  {\em Phys.Rev.} {\bf D66} (2002) 053005,
  [\href{http://xxx.lanl.gov/abs/hep-ph/0206211}{{\tt hep-ph/0206211}}].

\bibitem{lundberg2004weakly}
J.~Lundberg and J.~Edsj{\"o}, {\it {Weakly interacting massive particle
  diffusion in the solar system including solar depletion and its effect on
  Earth capture rates}},  {\em Phys.Rev.} {\bf D69} (2004) 123505,
  [\href{http://xxx.lanl.gov/abs/astro-ph/0401113}{{\tt astro-ph/0401113}}].

\bibitem{bertone2005particle}
G.~Bertone, D.~Hooper, and J.~Silk, {\it {Particle Dark Matter: Evidence,
  Candidates and Constraints}},  {\em Phys.Rept.} {\bf 405} (2005) 279--390,
  [\href{http://xxx.lanl.gov/abs/hep-ph/0404175}{{\tt hep-ph/0404175}}].

\bibitem{jungman1996supersymmetric}
G.~Jungman, M.~Kamionkowski, and K.~Griest, {\it {Supersymmetric Dark Matter}},
   {\em Phys.Rept.} {\bf 267} (1996) 195--373,
  [\href{http://xxx.lanl.gov/abs/hep-ph/9506380}{{\tt hep-ph/9506380}}].

\bibitem{cao2009dark}
Q.-H. Cao, E.~Ma, and G.~Shaughnessy, {\it {Dark Matter: The Leptonic
  Connection}},  {\em Phys.Lett.} {\bf B673} (2009) 152--155,
  [\href{http://xxx.lanl.gov/abs/0901.1334}{{\tt arXiv:0901.1334}}].

\bibitem{adriani2009anomalous}
{\bf PAMELA} Collaboration, O.~Adriani et~al., {\it {An Anomalous Positron
  Abundance in Cosmic Rays with Energies 1.5-100 GeV}},  {\em Nature} {\bf 458}
  (2009) 607--609, [\href{http://xxx.lanl.gov/abs/0810.4995}{{\tt
  arXiv:0810.4995}}].

\bibitem{adriani2010pamela}
{\bf PAMELA} Collaboration, O.~Adriani et~al., {\it {PAMELA Results on the
  Cosmic-Ray Antiproton Flux from 60 MeV to 180 GeV in Kinetic Energy}},  {\em
  Phys.Rev.Lett.} {\bf 105} (2010) 121101,
  [\href{http://xxx.lanl.gov/abs/1007.0821}{{\tt arXiv:1007.0821}}].

\bibitem{adriani2009new}
{\bf PAMELA} Collaboration, O.~Adriani et~al., {\it {A New Measurement of the
  Antiproton-to-Proton Flux Ratio up to 100 GeV in the Cosmic Radiation}},
  {\em Phys.Rev.Lett.} {\bf 102} (2009) 051101,
  [\href{http://xxx.lanl.gov/abs/0810.4994}{{\tt arXiv:0810.4994}}].

\bibitem{halzen2009indirect}
F.~Halzen and D.~Hooper, {\it {The Indirect Search for Dark Matter with
  IceCube}},  {\em New J.Phys.} {\bf 11} (2009) 105019,
  [\href{http://xxx.lanl.gov/abs/0910.4513}{{\tt arXiv:0910.4513}}].

\bibitem{abbasi2012design}
{\bf IceCube} Collaboration, R.~Abbasi et~al., {\it {The design and performance
  of IceCube DeepCore}},  {\em Astropart.Phys.} {\bf 35} (2012)
  [\href{http://xxx.lanl.gov/abs/1109.6096}{{\tt arXiv:1109.6096}}].

\bibitem{sjostrand2008brief}
T.~Sj{\"o}strand, S.~Mrenna, and P.~Z. Skands, {\it {A Brief Introduction to
  PYTHIA 8.1}},  {\em Comput.Phys.Commun.} {\bf 178} (2008) 852--867,
  [\href{http://xxx.lanl.gov/abs/0710.3820}{{\tt arXiv:0710.3820}}].

\bibitem{bell2011enhanced}
N.~Bell and K.~Petraki, {\it {Enhanced Neutrino Signals from Dark Matter
  Annihilation in the Sun via Metastable Mediators}},  {\em JCAP} {\bf 04}
  (2011) 003, [\href{http://xxx.lanl.gov/abs/1102.2958}{{\tt
  arXiv:1102.2958}}].

\bibitem{friedland2001evolution}
A.~Friedland, {\it {On the Evolution of the Neutrino State Inside the Sun}},
  {\em Phys.Rev.} {\bf D64} (2001) 013008,
  [\href{http://xxx.lanl.gov/abs/hep-ph/0010231}{{\tt hep-ph/0010231}}].

\bibitem{nakamura2010review}
{\bf Particle Data Group} Collaboration, K.~Nakamura et~al., {\it {Review of
  Particle Physics}},  {\em J.Phys.} {\bf G37} (2010) 075021.

\bibitem{abe2011indication}
{\bf T2K} Collaboration, K.~Abe et~al., {\it {Indication of Electron Neutrino
  Appearance from an Accelerator-produced Off-axis Muon Neutrino Beam}},  {\em
  Phys.Rev.Lett.} {\bf 107} (2011) 041801,
  [\href{http://xxx.lanl.gov/abs/1106.2822}{{\tt arXiv:1106.2822}}].

\bibitem{an2012observation}
{\bf Daya Bay} Collaboration, F.~An et~al., {\it {Observation of
  electron-antineutrino disappearance at Daya Bay}},
  \href{http://xxx.lanl.gov/abs/1203.1669}{{\tt arXiv:1203.1669}}.

\bibitem{wolfenstein1978neutrino}
L.~Wolfenstein, {\it {Neutrino Oscillations in Matter}},  {\em Phys.Rev.} {\bf
  D17} (1978) 2369--2374.

\bibitem{mikheev1985resonance}
S.~Mikheev and A.~Smirnov, {\it Resonanc{e enhancement of oscillations in
  matter and solar neutrino spectroscopy}},  {\em Sov. J. Nucl. Phys.(Engl.
  Transl.)} {\bf 42} (1985).

\bibitem{toshev1987exact}
S.~Toshev, {\it {Exact Analytical Solution of the Two Neutrino Evolution
  Equation in Matter with Exponentially Varying Density}},  {\em Phys.Lett.}
  {\bf B196} (1987) 170.

\bibitem{strumia2006neutrino}
A.~Strumia and F.~Vissani, {\it {Neutrino masses and mixings and...}},
  \href{http://xxx.lanl.gov/abs/hep-ph/0606054}{{\tt hep-ph/0606054}}.

\bibitem{lehnert2008neutrino}
R.~Lehnert and T.~J. Weiler, {\it {Neutrino Flavor Ratios as Diagnostic of
  Solar WIMP Annihilation}},  {\em Phys.Rev.} {\bf D77} (2008) 125004,
  [\href{http://xxx.lanl.gov/abs/0708.1035}{{\tt arXiv:0708.1035}}].

\bibitem{mikheev1987resonance}
S.~Mikheev and A.~Smirnov, {\it {Resonance oscillations of neutrinos in
  matter}},  {\em Sov.Phys.Usp.} {\bf 30} (1987) 759.

\bibitem{kachelriess2001nonadiabatic}
M.~Kachelriess and R.~Tomas, {\it {Nonadiabatic level crossing in resonant and
  nonresonant neutrino oscillations}},  {\em Phys.Rev.} {\bf D64} (2001), no.~7
  073002.

\bibitem{gandhi1996ultrahigh}
R.~Gandhi, C.~Quigg, M.~H. Reno, and I.~Sarcevic, {\it {Ultrahigh-Energy
  Neutrino Interactions}},  {\em Astropart.Phys.} {\bf 5} (1996) 81--110,
  [\href{http://xxx.lanl.gov/abs/hep-ph/9512364}{{\tt hep-ph/9512364}}].

\bibitem{dutta2001propagation}
S.~Dutta, M.~Reno, I.~Sarcevic, and D.~Seckel, {\it {Propagation of Muons and
  Taus at High Energies}},  {\em Phys.Rev.} {\bf D63} (2001) 094020,
  [\href{http://xxx.lanl.gov/abs/hep-ph/0012350}{{\tt hep-ph/0012350}}].

\bibitem{beacom2003measuring}
J.~F. Beacom, N.~F. Bell, D.~Hooper, S.~Pakvasa, and T.~J. Weiler, {\it
  {Measuring Flavor Ratios of High-Energy Astrophysical Neutrinos}},  {\em
  Phys.Rev.} {\bf D68} (2003) 093005,
  [\href{http://xxx.lanl.gov/abs/hep-ph/0307025}{{\tt hep-ph/0307025}}].

\bibitem{ahrens2004sensitivity}
{\bf IceCube} Collaboration, J.~Ahrens et~al., {\it {Sensitivity of the IceCube
  Detector to Astrophysical Sources of High Energy Muon Neutrinos}},  {\em
  Astropart.Phys.} {\bf 20} (2004) 507--532,
  [\href{http://xxx.lanl.gov/abs/astro-ph/0305196}{{\tt astro-ph/0305196}}].

\bibitem{abbasi2012multiyear}
{\bf IceCube} Collaboration, R.~Abbasi et~al., {\it {Multi-year search for dark
  matter annihilations in the Sun with the AMANDA-II and IceCube detectors}},
  {\em Phys. Rev.} {\bf D85} (2012) 042002,
  [\href{http://xxx.lanl.gov/abs/1112.1840}{{\tt arXiv:1112.1840}}].

\bibitem{fukushima2012brem}
K.~Fukushima, Y.~Gao, J.~Kumar, and D.~Marfatia, {\it {Bremsstrahlung
  signatures of dark matter annihilation in the Sun}},
  \href{http://xxx.lanl.gov/abs/1208.1010}{{\tt arXiv:1208.1010}}.

\bibitem{steigman2012precise}
G.~Steigman, B.~Dasgupta, and J.~F. Beacom, {\it {Precise relic WIMP abundance
  and its impact on searches for dark matter annihilation}},  {\em Phys. Rev.}
  {\bf D86} (2012) 023506, [\href{http://xxx.lanl.gov/abs/1204.3622}{{\tt
  arXiv:1204.3622}}].

\end{thebibliography}\endgroup

\end{document}